\documentstyle[11pt,aaspp4,epsf]{article}

\newcommand{\simgt}{\lower.5ex\hbox{$\; \buildrel > \over \sim \;$}}
\newcommand{\simlt}{\lower.5ex\hbox{$\; \buildrel < \over \sim \;$}}

\newcommand{\vect}[1]{\mbox{\boldmath$#1$}}
\def\pp{\par\parshape 2 0truecm 15.5truecm 1truecm 14.5truecm\noindent}
\def\II{{\rm I\hspace{-0.5mm}I}}
\def\III{{\rm I\hspace{-0.5mm}I\hspace{-0.5mm}I}}

\def\LC{{\rm R}}
\def\RS{{\rm S}}
\def\PLC{{P^{\rm LC}_{\rm S}}}
\def\xiLC{{\xi^{\rm LC}_{\rm S}}}

\begin{document}
\parbox{\hsize}{
\begin{flushright}
HUPD-9911~~~~~~~~~~~\\
August 1999~~~~~~~~~~~\\
Revised October 1999~~~~~~~~~~~\\
\end{flushright}}

\title{Redshift-space Distortions of the Power Spectrum \\
of Cosmological Objects on a Light Cone~:\\
Explicit Formulations and Theoretical Implications}

\bigskip
\author{Hiroaki {\small NISHIOKA} and Kazuhiro {\small YAMAMOTO}
\\
{\it Department of Physics, Hiroshima
    University, Higashi-Hiroshima 739-8526, Japan}}

\bigskip
\received{1999 August }
\accepted{1999 ??}
\begin{abstract}
We examine the effects of the linear and the cosmological 
redshift-space distortions on the power spectrum of 
cosmological objects on a light cone. We develop theoretical 
formulae for the power spectrum in linear theory of density perturbations
in a rigorous manner starting from first principle corresponding 
to Fourier analysis. Approximate formulae, which are useful properly
to incorporate the redshift-space distortion effects into the power 
spectrum are derived, and the validity is examined.
Applying our formulae to galaxy and quasar samples which roughly 
match the SDSS survey, we will show how the redshift-space distortions
distort the power spectrum on the light cone quantitatively.
\end{abstract}

\keywords{ cosmology: theory - dark matter - large-scale structure of
universe -- galaxies: distances and redshifts -- quasars: general}

\section{INTRODUCTION}
The wide-field redshift surveys of galaxies and quasars, like the 
Two-degree Field (2dF) and the Sloan Digital Sky Survey (SDSS), 
are key projects in the area of observational cosmology. 
These surveys provide three dimensional maps of the distribution 
of cosmological objects extended to high redshifts. The spatial 
distribution of the objects depends on various properties
of the universe such as, initial density fluctuations, formation 
process of the cosmological objects, bias mechanism, 
luminosity function, cosmological parameters, and dark 
matter. The comparison of the observational data with 
theoretical predictions will put serious constraints on 
the cosmological models. Especially the precise measurement
of the power spectrum from the galaxy surveys is a promising 
project for the cosmological parameters and the initial density 
fluctuations, as well as the measurement of the cosmic microwave 
background anisotropies.

Two-point correlation function and the power spectrum are 
the most used tools to quantify the clustering of 
the cosmological objects. 
From a theoretical point of view, these statistical quantities
that are well defined on the constant-time hypersurface have been 
used so far. However, cosmological observations are carried out 
only on a light cone. This fact requires statistical quantities which 
properly incorporate the light-cone effect. For the 2dF and 
SDSS surveys the light-cone effect becomes important because of the 
depth of the samples. This importance of the light-cone 
effect was discussed by several authors 
(Matarrese et~al. 1997; Matsubara, Suto, \& Szapudi 1997; Nakamura, 
Matsubara, \& Suto 1998; de Laix \& Starkman 1998; Moscardini et~al. 1998).
Recently, a rigorous framework to incorporate the light-cone effect
into the two-point correlation function and the power spectrum
has been developed (Yamamoto \& Suto 1999, Papar I; Nishioka \& Yamamoto 
1999, Paper \II; Suto et~al. 1999 ; Yamamoto, Nishioka, \& Suto 1999, 
Paper \III). 

The other important observational effect is the redshift-space distortion.
Because cosmological surveys are carried out in redshift space, 
the peculiar velocity of sources distorts the distribution of the 
cosmological objects. This effect has been extensively investigated 
in linear theory (Davis \& Peebles 1983; Kaiser 1987; Hamilton 1998;
Szalay, Matsubara, \& Landy 1998) 
and also in a nonlinear regime (Cole, Fisher, \& Weinberg 1994; 
Suto, et~al 1999; Magira, Jing, \& Suto 1999).  
It has been also pointed out that the geometrical effect 
originating from the expansion of the universe causes another 
apparent distortion in the distribution of cosmological objects,
i.e., cosmological redshift-space distortion 
(Alcock \& Paczynski 1979; Ryden 1995; Matsubara \& Suto 1996; 
Ballinger, Peacock, \& Heavens 1996; Suto, et~al 1999; Magira, 
Jing, \& Suto 1999; Matsubara 1999). Though the authors in these 
previous works 
have focused on determining the cosmological parameters by using 
the cosmological redshift-space distortion,
we note that the cosmological redshift-space distortion gives 
rise to a problem in determining the power spectrum because the 
cosmological parameters of the universe have not been well established.

In this paper we focus on the power spectrum on a light cone, 
taking into account the redshift-space distortion effects.
We derive formulae for the power spectrum in a rigorous manner
and examine how these effects distort the power spectrum 
measured in wide- and deep-field surveys. 
We adopt the following strategy. We first define an estimator 
for the power spectrum which can be computed from a sample of
cosmological sources on a light cone, according to the conventional 
Fourier analysis. Then we calculate the ensemble average of the
estimator, which we define the power spectrum on a light cone. 
Comparing this with the conventional power spectrum defined on a 
constant-time hypersurface in real space, we show how the 
distortion appears in the shape of the power spectrum 
on the light cone quantitatively, modeling galaxy and quasar 
samples which roughly match the SDSS samples.

In the present paper we work in linear theory, which enables us
to derive formulae in a rigorous manner. However, this limits 
the validity of our results only to the large scales because
the nonlinear effect affects the clustering features
on the small scales. So the extension to include the nonlinear 
effect is needed, and a plausible extension is discussed in Paper \III.

This paper is organized as follows: In \S 2 we define the statistical
estimators of the power spectrum and the two-point correlation function
on a light cone.
We will find that a familiar conventional relation between the power 
spectrum and the two-point correlation function, which is defined 
on a constant-time hypersurface, holds even for those thus defined 
on the light cone.
In \S 3 we consider the linear redshift-space distortion effect
on the power spectrum. Some parts in this section have been discussed 
in Paper \III. However, the approximate formula to 
incorporate the linear redshift-space distortion,
which was introduced in Paper \III~ in an intuitive manner, 
is derived in the rigorous manner in the present section.
And the validity of the approximate formula is 
examined in detail assuming galaxy and quasar samples. 
In \S 4 we consider the cosmological redshift-space distortion effect.
Then the rigorous and the approximate formulae for the power spectrum 
on the light cone are derived. Then we examine the validity of 
approximation, and discuss the theoretical implications.
\S 5 is devoted to summary and conclusions. 
Throughout this paper we use the unit in which the light 
velocity $c$ equals 1. 

For definiteness, we adopt notations similar to those in Paper \III, 
and use a superscript LC to indicate quantities on the light cone 
explicitly. The power spectrum without the superscript $P(k)$ 
denotes the CDM power spectrum defined on a constant-time 
hypersurface at present. And we use subscripts (superscripts) 
R and S to indicate quantities in real and redshift spaces, 
respectively.

\section{DEFINITION OF THE POWER SPECTRUM}
In this section we first define statistical estimators for the
two-point correlation function and the power spectrum of cosmological 
objects on a light-cone hypersurface 
defined by an observer in redshift space.
We define the statistical estimators for the two-point 
correlation function and the power spectrum, independently, 
which can be computed when a set of a survey catalog is obtained. 
Then we will show that a familiar fundamental relation holds between 
thus defined two-point correlation function and the power spectrum.

In a redshift survey, a position of a cosmological object is 
specified by the redshift $z$ and the direction (unit) vector 
$\vect{\gamma}$.
Cosmological distance to the object is evaluated 
from the redshift $z$ through some converting formula. And a map 
of the cosmological objects is obtained. The statistical 
quantities, the power spectrum and the two-point correlation 
function, are computed from the map. 
By introducing the variable $s$ to specify the cosmological 
distance (the radial coordinate) in redshift space, we 
write the number density field of the cosmological objects in 
redshift space $n^\RS(s,\vect{\gamma})$ and the mean number density 
${\tilde n}^\RS(s)$. \footnote{We assume infinitely thin binwidth for
the 'observed' data set.}
Following the conventional treatment of the power spectrum
(e.g., Feldman, Kaiser, \& Peacock 1994), we introduce the number 
density field
\begin{equation}
  F(s,\vect{\gamma})={n^\RS(s,\vect{\gamma})-n^\RS_{\rm syn}(s,\vect{\gamma})
  \over\biggl[{\displaystyle \int d^3 \vect{s} {\tilde n}^\RS(s)^2}\biggr]
^{1\over2}},
\label{eq1}
\end{equation}
where $n_{\rm syn}^\RS(s,\vect{\gamma})$ is the number density field 
of objects for a synthetic catalog without structure which 
has the same mean number density ${\tilde n}^\RS(s)$ as that of the 
cosmological objects $n^\RS(s,\vect{\gamma})$.
To be specific for the synthetic catalog, if we write 
$n^\RS(s,\vect{\gamma})$ and $n_{\rm syn}^\RS(s,\vect{\gamma})$ as
\begin{eqnarray}
  n^\RS(s,\vect{\gamma})&=&{\tilde n}^\RS(s)[1+\Delta^\RS(s,\vect{\gamma})], 
\label{eq2} \\
  n^\RS_{\rm syn}(s,\vect{\gamma})&=&{\tilde n}^\RS(s)
[1+\Delta^\RS_{\rm syn}(s,\vect{\gamma})],
\label{eq3}
\end{eqnarray}
respectively, we write
\begin{equation}
  \Bigl<\Delta^\RS_{\rm syn}(s_1,\vect{\gamma}_1)
\Delta^\RS_{\rm syn}(s_2,\vect{\gamma}_2)\Bigr>
  =\Bigl<\Delta^\RS(s_1,\vect{\gamma}_1)\Delta^\RS_{\rm syn}
(s_2,\vect{\gamma}_2)  \Bigl> =0.
\label{eq4}
\end{equation}

When the density field $F(s,\vect{\gamma})$ is given, one may 
compute the following two-point statistics, 
\begin{equation}
  \xi^{obs}(R)=\int {d\Omega_{\hat{\vect{R}}}\over 4\pi}
  \int d^3 \vect{s}_1
  \int d^3 \vect{s}_2 F(s_1,\vect{\gamma}_1)F(s_2,\vect{\gamma}_2)
  \delta^{(3)}(\vect{s}_1-\vect{s}_2-\vect{R}),
\label{eq8}
\end{equation} 
where $\vect{s}_1=s_1\vect{\gamma}_1$ and $\vect{s}_2=s_2\vect{\gamma}_2$
and $R=\vert\vect{R}\vert$, and $\hat{\vect{R}}=\vect{R} /R$. 
Note that this definition of the estimator for the two-point 
correlation function is slightly different from that in the 
previous paper (Paper I; Paper \II). However, we will show that the 
difference is not practical and that the same formula is derived in 
the limit of the distant observer approximation (equation [\ref{eq68.5}]).

Here $\xi^{obs}(R)$ is a mathematical expression for the two-point
correlation function computed from a conventional data processing
for a set of data $n^\RS(s,\vect{\gamma})$ (for an observer). 
As is well-known as the problem 
of the cosmic variance, we can only predict the ensemble average of the
statistical estimator, where the ensemble average means to
average the estimator over many universes for different observers.
Assuming that the selection function does not depend on the direction
$\vect{\gamma}$, the ensemble average of the estimator is written as
\begin{equation}
  \xiLC(R)=\Bigl<\xi^{obs}(R)\Bigr>=\int {d\Omega_{\hat{\vect{R}}}\over 4\pi}
  \int d^3\vect{s}_1\int d^3\vect{s}_2
  \Bigl<F(s_1,\vect{\gamma}_1)F(s_2,\vect{\gamma}_2)
  \delta^{(3)}(\vect{s}_1-\vect{s}_2-\vect{R})\Bigr>,
\label{eq9}
\end{equation}
which we define as the two-point correlation function on a light cone 
in redshift space.

Next we consider the power spectrum. 
When the density field $F(s,\vect{\gamma})$ is given, one can
compute the Fourier coefficient of equation ($\ref{eq1}$)
\begin{equation}
  {\cal{F}}(\vect{k})
  =\int d^3 \vect{s}F(s,\vect{\gamma})e^{i\vect{k}\cdot \vect{s}}
  ={{\displaystyle \int d^3 \vect{s}[n^\RS(s,\vect{\gamma})
  -n^\RS_{\rm syn}(s,\vect{\gamma})]e^{i\vect{k}\cdot \vect{s}}}
  \over {\displaystyle \biggl[\int d^3 \vect{s} {\tilde n}^\RS(s)^2\biggr]^{1\over 2}}},
\label{eq6}
\end{equation}
and the power spectrum may be computed as follows (Feldman, Kaiser, 
\& Peacock 1994),
\begin{equation}
  P^{obs}(k)=\int{d \Omega_{\hat{\vect{k}}}\over 4\pi}
  \vert{\cal{F}}(\vect{k})\vert^2,
\label{eq5}
\end{equation} 
where $\vect{k}$ denotes the wavenumber vector,
$k=\vert\vect{k}\vert$, and $\hat{\vect{k}}=\vect{k}/k$. 
Similar to the case of the two-point correlation function, 
$P^{obs}(k)$ models a conventional estimation of the power 
spectrum computed from a set of data $F(s,\vect{\gamma})$ for an observer. 
The ensemble average of the statistical estimator is written as
\begin{equation}
  \PLC(k)=\Bigl<P^{obs}(k)\Bigr>=\int {d\Omega_{\hat{\vect{k}}}\over 4\pi}
  \Bigl<\vert{\cal{F}}(\vect{k})\vert^2\Bigr>,
\label{eq7}
\end{equation}
which we define as the power spectrum on a light cone 
in redshift space 
(see also Paper \III).

When we adopt the above definitions, we show the 
familiar fundamental relation which holds between the 
two-point correlation function and the power spectrum, as follows.
With the use of the relation
\begin{equation}
  \delta^{(3)}(\vect{s}_1-\vect{s}_2-\vect{R})
  =\int{d^3\vect{k}\over(2\pi)^3}
  e^{i\vect{k}\cdot(\vect{s}_1-\vect{s}_2-\vect{R})},
\end{equation} 
equation ($\ref{eq9}$) is written as 
\begin{eqnarray}
  \xiLC(R)&=&\int {d\Omega_{\hat{\vect{R}}}\over 4\pi}
  \int{d^3\vect{k}\over(2\pi)^3}
  \Bigl<|F(\vect{k})|^2\Bigr>e^{-i\vect{k}\cdot \vect{R}}
\nonumber 
\label{eq11} \\ 
  &=&{1\over 2\pi^2}\int dk k^2 \PLC(k)j_0(kR).
\label{eq12}
\end{eqnarray}
The inverse Fourier transformation yields,
\begin{eqnarray}
  \PLC(k) = 4\pi \int dR R^2 \xiLC(R)j_0(kR).
\label{eq13}
\end{eqnarray}
Equations (\ref{eq12}) and (\ref{eq13}) are the familiar formulae 
which hold for the conventional two-point correlation function 
and the power spectrum defined on a constant-time hypersurface.

\section{LINEAR REDSHIFT-SPACE DISTORTION}
The distortion due to the linear peculiar velocity field of cosmological 
objects is called as the linear redshift-space distortion. 
In this section we consider the linear redshift-space distortion
under the assumption that the cosmological model is exactly determined.
An error of assuming the cosmological model causes 
another distortion, i.e., the cosmological redshift-space 
distortion, which will be investigated in the next section.

\subsection{Two-point correlation function on a light cone}
In this subsection we first calculate the two-point correlation function,
taking into account the linear redshift-space distortion. Formulation
is developed within linear theory of density perturbations based on the 
CDM scenario. Because the definition of the two-point correlation 
function is different from that in the previous paper (Papers \II), 
derivation of the formula is slightly different from the previous one. 
We briefly summarize the calculation for definiteness.

Throughout the present paper, we consider the spatially-flat 
Friedmann-Lemaitre universe, whose line element is expressed as
\begin{equation}                    
  ds^2=a^2(\eta)\Bigl[-d\eta^2+d\chi^2+\chi^2 d\Omega^2_{(2)}\Bigr],
\label{eq14}
\end{equation} 
where $\eta$  is the conformal time, $a$ is the scale factor, 
$\chi$ is the radial coordinate, and $d\Omega_{(2)}^2$ is the line
element of the unit two-sphere. We normalize the scale factor to be 
unity at present, $a(\eta_0)=1$. Then the Friedmann equation is 
\begin{equation}
  \biggl({{\dot a}\over a}\biggr)^2=H_0^2\biggl(
  {\Omega_0\over a}+a^2\Omega_{\lambda}\biggr),
\label{eq15}
\end{equation}
where $H_0=100h{\rm km/s/Mpc}$ is the Hubble parameter,
$\Omega_0(=1-\Omega_\lambda)$ is the density parameter,
and the dot denotes $\eta$-differentiation.

Since we locate a fiducial observer at the origin of coordinates
($\eta=\eta_0,\chi=0$), a cosmological object at $\chi$ and $\eta$
on the light cone hypersurface of the observer satisfies the simple 
relation $\eta=\eta_0-\chi$. 
We introduce the radial coordinate $r$ to express the three 
dimensional (real) space on the light-cone hypersurface, in which the 
position of a source is specified by $(r,\vect{\gamma})$, where 
$\vect{\gamma}$ is an unit vector along the line of sight. 
Essentially, $r$ is equivalent to $\chi$, however, we use 
$r$ for mathematical conveniences and to represent that 
quantities are defined on the light cone explicitly.
Then we denote the metric of the three dimensional space on the light cone as, 
\begin{equation}
  ds^2_{\rm LC}=dr^2+r^2 d \Omega^2_{(2)},
\label{eq16}
\end{equation}
and we write the number density of sources on the light cone as 
$n^{\LC}(r,\vect{\gamma})$ , which is simply related to the comoving
number density of objects at a conformal time $\eta$ and at a position,
$(\chi,\vect{\gamma})$ , $n(\eta,\chi,\vect{\gamma})$ as
\begin{equation}
  n^{\LC}(r,\vect{\gamma})=n(\eta,\chi,\vect{\gamma})
  |_{\eta\rightarrow\eta_0 -r, \chi\rightarrow r}.
\label{eq17}
\end{equation}
By using the mean observed comoving number density ${\tilde n}(\eta)$ at
time $\eta$ and the density fluctuation of luminous objects 
$\Delta(\eta,\chi,\vect{\gamma})$, we write
\begin{equation}
n(\eta,\chi,\vect{\gamma})={\tilde n}(\eta)[1+\Delta(\eta,\chi,\vect{\gamma})].
\label{eq18}
\end{equation}
Then equation ($\ref{eq17}$) is rewritten as
\begin{equation}
n^{\LC}(r,\vect{\gamma})={\tilde n}^{\LC}(r)[1+\Delta^{\LC}
(r,\vect{\gamma})],
\label{eq19}
\end{equation}
where we defined
\begin{eqnarray}
  &&{\tilde n}^{\LC}(r)\equiv {\tilde n}(\eta)|_{\eta\rightarrow\eta_0-r},
  \hspace{5mm}
  \Delta^{\LC}(r,\vect{\gamma})\equiv \Delta(\eta,\chi,\vect{\gamma})
  |_{\eta\rightarrow\eta_0-r, \chi\rightarrow r}.
\label{eq21}
\end{eqnarray}
Note that the mean observed number density ${\tilde n}(\eta)$ is different from the 
mean number density of the objects at $\eta$ by a factor of a selection 
function.
%

Now we consider the two-point correlation function ($\ref{eq9}$)
in redshift space. The relation between the redshift space 
and the real space is specified as follows,
\begin{eqnarray}
  &&n^\RS(s,\vect{\gamma}) s^2 ds d\Omega_{\vect{\gamma}} 
  =n^{\LC}(r,\vect{\gamma}) r^2 dr d\Omega_{\vect{\gamma}},
\label{eq23}
\\
  &&n^\RS_{\rm syn}(s,\vect{\gamma})s^2 ds  d\Omega_{\vect{\gamma}}
  =n^{\LC}_{\rm syn}(r,\vect{\gamma}) r^2 dr d\Omega_{\vect{\gamma}},
\label{eq28}
\end{eqnarray}
and
\begin{equation}
s=r+\delta r(r,\vect{\gamma}),
\label{eq24}
\end{equation}
where $\delta r(r,\vect{\gamma})$ in equation ($\ref{eq24}$) represents 
the apparent shift in the comoving radial coordinate due to peculiar 
velocity. $\delta r(r,\vect{\gamma})$ is of order of the velocity 
perturbations, and the explicit expression is summarized in 
Appendix. Equation (\ref{eq28}) defines the number density of the 
synthetic catalog in real space $n^{\LC}_{\rm syn}(r,\vect{\gamma})$.

With the use of the above equations we obtain the following expression
from equation (\ref{eq9}),
\begin{eqnarray}
  \xiLC(R)&=&\biggl[\int d^3 \vect{r} {\tilde n}^{\LC}(r)^2\biggr]^{-1}
  \int {d \Omega_{\hat{\bf R}}\over 4\pi}
  \int dr_1 r_1^2 d\Omega_{\vect{\gamma}_1} 
  \int dr_2 r_2^2 d\Omega_{\vect{\gamma}_2} 
\nonumber 
\\
&&\times\Bigl<[n^{\LC}(r_1,\vect{\gamma}_1)-
  n^{\LC}_{\rm syn}(r_1,\vect{\gamma}_1)]
  [n^{\LC}(r_2,\vect{\gamma}_2)-n^{\LC}_{\rm syn}(r_2,\vect{\gamma}_2)]
\nonumber 
\\
  &&\times\delta^{(3)}(\vect{r}_1+\delta \vect{r}_1
  -\vect{r}_2-\delta \vect{r}_2-\vect{R})\Bigr>,
\label{eq27}
\end{eqnarray}
where $\vect{r} +\delta \vect{r}=(r+\delta r)\vect{\gamma}$, and
we have assumed
\begin{equation}
  \int d^3 \vect{s} {\tilde n}^\RS(s)^2 = \int d^3 \vect{r} {\tilde n}^{\LC}(r)^2.
\label{eq29}
\end{equation}
%
Note that the number density field of the synthetic catalog in 
redshift space $n^\RS_{\rm syn}(s,\vect{\gamma})$ has no structure. 
Then the number density field in real space 
$n^{\LC}_{\rm syn}(r,\vect{\gamma})$, which is defined by equation 
($\ref{eq28}$), has a structure. Writing the number density field as
\begin{equation}
  n_{\rm syn}^{\LC}(r,\vect{\gamma})={\tilde n}^{\LC}(r)[1+\Delta_{\rm syn}^{\LC}
  (r,\vect{\gamma})],
\label{eq30}
\end{equation}
the density contrast $\Delta_{\rm syn}^{\LC}(r,\vect{\gamma})$
is expressed as (Hamilton 1998),
\begin{equation}
\Delta^{\LC}_{\rm syn}(r,\vect{\gamma})=\Delta^\RS_{\rm syn}(r,\vect{\gamma})
+\biggl({\partial\over \partial r}+{\kappa(r)\over r}\biggl)
\delta r(r,\vect{\gamma}),
\label{eq31}
\end{equation}
where
\begin{equation}
\kappa(r)={\partial \ln r^2 {\tilde n}^{\LC}(r)\over \partial \ln r},
\label{eq32}
\end{equation}
and we used $\Delta^\RS_{\rm syn}(r,\vect{\gamma})
=\Delta^\RS_{\rm syn}(s,\vect{\gamma})$,
which holds within the linear order of perturbations.
In the leading order of the perturbative expansion, 
equation (\ref{eq27}) is rewritten as  
\begin{eqnarray}
  \xiLC(R)&=&
  \biggl[\int d^3 \vect{r}{\tilde n}^{\LC}(r)^2\biggr]^{-1}\int 
  {d\Omega_{\hat{\vect{R}}}\over 4\pi}\int dr_1r_1^2 d\Omega_{\vect{\gamma}_1}
  \int dr_2r_2^2 d\Omega_{\vect{\gamma}_2}
  {\tilde n}^{\LC}(r_1){\tilde n}^{\LC}(r_2)
\nonumber \\
  &&\times\Bigl<[\Delta^{\LC}(r_1,\vect{\gamma}_1)
  -\biggl({\partial\over\partial r_1}+{\kappa(r_1)\over r_1}\biggr)
  \delta r_1(r_1,\vect{\gamma}_1)]
\nonumber \\
&&\times[\Delta^{\LC}(r_2,\vect{\gamma}_2)
  -\biggl({\partial\over\partial r_2}+{\kappa(r_2)\over r_2}\biggr)
  \delta r_2(r_2,\vect{\gamma}_2)]\Big>
  \delta^{(3)}(\vect{r}_1-\vect{r}_2-\vect{R})~,
\label{eq34}
\end{eqnarray}
where we used equations (\ref{eq19}), (\ref{eq30}), (\ref{eq31}), and 
(\ref{eq4}).

After straightforward calculations based on the CDM model with the 
linear biasing (equation [\ref{eq50}]), we obtain the following expression
(see Appendix),
\begin{eqnarray}
\xiLC(R)&=&\biggl[\int dr r^2{\tilde n}^{\LC}(r)^2\biggr]^{-1}
{1\over 2R}\int\!\!\!\int_{\cal S}dr_1dr_2r_1r_2
{\tilde n}^{\LC}(r_1){\tilde n}^{\LC}(r_2)
\nonumber \\
&&\times{1\over 2\pi^2}\int dk k^2 P(k)
\prod^2_{i=1}\biggl[b(k;\eta_0-r_i)D_1(\eta_0-r_i)\biggr]
\nonumber \\
&&\times\biggl[j_0(kR)+\beta(k;\eta_0-r_2)I(k;R;r_1,r_2)
+\beta(k;\eta_0-r_1)I(k;R;r_2,r_1)\biggr.
\nonumber \\
&&\biggl.
+\beta(k;\eta_0-r_1)\beta(k;\eta_0-r_2)J(k;R;r_1,r_2)\biggr],
\label{eq68.5}
\end{eqnarray}
where $\cal S$ denotes the region $|r_1-r_2|\leq R\leq r_1+r_2$, 
$P(k)$ is the CDM power spectrum at present, $D_1(\eta)$ is the
linear growth rate normalized to be unity at present,
$I(k;R;r_1,r_2)$ and $J(k;R;r_1,r_2)$ are defined by 
equations ($\ref{eq69}$) and ($\ref{eq70}$), respectively,
and $\beta(k;\eta)$ is defined as
\begin{equation}
\beta(k;\eta)
={1\over b(k;\eta)}{d \ln D_1(\eta)\over d \ln a(\eta)}.
\label{eq71}
\end{equation}

\subsection{Distant Observer Approximation}
As is shown in Paper \II, equation (\ref{eq68.5})
reduces to a simple form by applying the plane-parallel, or 
distant observer, approximation. In the case $R\ll 2r_{\rm max}$,
where $2r_{\rm max}$ is the size of the survey volume, we can use 
the approximation
\begin{equation}
\int\!\!\!\int_{\cal S}dr_1dr_2\simeq
\int dr_1\int^R_{-R}dx
\label{eq72}
\end{equation} 
where $x=r_2-r_1$, and $I(k;R;r_1,r_2)$ and $J(k;R;r_1,r_2)$ 
are approximated as
\begin{eqnarray}
I(k;R;r_1,r_2)&=&{j_1(kR)\over kR}-{j_2(kR)\over R^2}x^2, \\
\label{eq73}
J(k;R;r_1,r_2)&=&3{j_2(kR)\over (kR)^2}-6{j_3(kR)\over kR^3}x^2
+{j_4(kR)\over R^4}x^4,
\label{eq74}
\end{eqnarray}
respectively. Integration over $x$ yields,  
\begin{equation}
  \xiLC(R)\simeq
 {
   {\displaystyle
    \int dr r^2 {\tilde n}^{\LC}(r)^2 
    \xi[R,z(r)]_{\rm source}
    }
\over
    {\displaystyle
    \int dr r^2 {\tilde n}^{\LC}(r)^2
    }
} ,
\label{xisapprox}
\end{equation}
where we defined
\begin{eqnarray}
  \xi[R,z(r)]_{\rm source}&=&
  {1\over 2\pi^2}\int dk k^2 P(k)D_1(\eta_0-r)^2 b(k;\eta_0-r)^2
\nonumber
\\
  &&~~~~\times\biggl[1+{2\over 3}\beta(k;\eta_0-r)+{1\over 5}\beta(k;\eta_0-r)^2
  \biggr]j_0(kR).
\label{eq76}
\end{eqnarray}
This formula indicates that the two-point correlation function
on a light cone is obtained by averaging the correlation function
at each time defined on a constant-time hypersurface 
$\xi[R,z(r)]_{\rm source}$ by weighting ${\tilde n}^{\LC}(r)^2$.
Note that this formula has been rigorously derived only 
in the framework of linear theory of density perturbations.

\subsection{Power spectrum on a light cone} 
The power spectrum is obtained from the two-point correlation
function as described in section 2.
By substituting equation (\ref{eq68.5}) into (\ref{eq13}) we 
compute the power spectrum  on the light cone $\PLC(k)$. Some 
aspects of the power spectrum on the light cone have been discussed 
in Paper \III. Although a different method was adopted to 
compute the power spectrum in Paper \III, we find that 
the approach employed in this paper is rather useful.
We derive the simple expression for the power 
spectrum, which was used without detailed verification in Paper \III,
and discuss the validity of approximation in subsection 3.5.

The approximate formula for the power spectrum is easily obtained
by substituting equation ($\ref{xisapprox}$) into ($\ref{eq13}$), 
\begin{equation}
\PLC(k)\simeq\alpha(k)P(k),
\label{eq80}
\end{equation}
where
\begin{equation}
  \alpha(k)=
 {
   {\displaystyle
    \int dr r^2 {\tilde n}^{\LC}(r)^2
    b(k;\eta_0-r)^2D_1(\eta_0-r)^2 
    \biggl[1+{2\over3}\beta(k;\eta_0-r)+{1\over5}\beta(k;\eta_0-r)^2
    \biggr]}
\over
    {\displaystyle
    \int dr r^2 {\tilde n}^{\LC}(r)^2
    }
} .
\label{defalpha}
\end{equation}
This equation justifies the extended formula (20) in Paper \III. 

In the case that the bias does not depend on scales of density 
fluctuation or $k$, i.e., $b(k;\eta)=b(\eta)$,  $\alpha(k)$ becomes a
constant $\alpha$. Thus in this case the light-cone effect and the
linear redshift-space distortion are described by the constant
$\alpha$. In the small scales of large wavenumber of $k$, the
nonlinearity distorts the power spectrum in redshift space, 
and produces an additional scale-dependence in the power spectrum 
on the light-cone. Importance of the nonlinearity in small scales 
is extensively discussed in Paper \III~(see also Magira, Jing, \& Suto 1999). 

\subsection{Model of samples}
%
In subsection 3.5 we apply the formulae developed in the previous 
subsections to galaxy and quasar samples, and discuss the validity 
of approximation by comparing the approximate formula with the exact 
formula using numerical calculations.
\footnote{
The terminology, 'exact' formula, means the exact expression rigorously
derived from definition ($\ref{eq9}$) or ($\ref{eq7}$) within the 
linear theory of density perturbation and bias.
And the 'approximate' formula means its version in the distant 
observer limit. In a realistic situation, the nonlinearity of
density perturbations and complex features of bias cause 
additional deformation of the power spectrum. 
(See also Paper \III, in which some
aspects of the nonlinear effects are discussed.)
}
For that purpose, we need to model galaxy and quasar samples 
and bias. For a galaxy luminosity function, we adopt a B-band 
luminosity function of the APM galaxies (Loveday et~al. 1992) 
fitted to Schechter function
\begin{equation}
\phi(L)dL=\phi^*\biggl({L\over L^*}\biggr)^\nu
\exp \biggl(-{L\over L^*}\biggr)d\biggl({L\over L^*}\biggr),
\end{equation}  
with $\phi^*=1.40\times 10^{-2}h^3{\rm Mpc}^{-3}$, $\nu=-0.97$,
and $M_B^*=-19.50+5\log_{10}h$. Then the comoving number density of
galaxies at $z$ which are brighter than the limiting magnitude
$B_{\rm lim}$ is given by
\begin{equation}
{\tilde n}(z,<B_{\rm lim})=\int^{\infty}_{L(B_{\rm lim},z)}\phi(L)dL
=\phi^*\Gamma[\nu+1,x(B_{\rm lim},z)],
\end{equation}
where
\begin{equation}
  x(B_{\rm lim},z)=
  {L(B_{\rm lim},z)\over L^*}
  =\biggl[{d_L(z)\over 1h^{-1}{\rm Mpc}}\biggr]^2 10^{2.2-0.4B_{\rm lim}},
\end{equation}
and $\Gamma[\nu,x]$ is the incomplete Gamma function.
We consider the galaxy sample in the range of redshift 
$0\leq z\leq 0.2$, and adopt the B-band limiting magnitude 
19 to match the SDSS spectroscopic sample.

We also assume quasar samples which roughly match the SDSS quasar 
survey, where we use a selection function for B-band magnitude limited 
samples on the basis of the luminosity function by 
Wallington \& Narayan (1993) and adopt the B-band limiting
magnitude 20 (see also Nakamura \& Suto 1997; Paper I). 
And we assume the depth of the survey volume, $z_{\rm max}=5$.

As for the bias, we adopt the simple scale-independent
model (Paper \III), 
which is a phenomenological extension of the Fry's bias model (1996), 
\begin{equation}
b(\eta)=1+{1\over [D_1(\eta)]^p}(b_0-1),
\label{bias}
\end{equation}
where $b_0=b(\eta_0)$ is the bias parameter at present, and
$p$ is the constant which controls time-evolution.
Note that this model reduces to the case of no bias when
the constant $b_0$ is unity and that the case $p=1$ is
equivalent to the model by Fry (1996).
  
We consider the SCDM(standard cold dark matter) and the LCDM(Lambda cold
dark matter) models, which have $(\Omega_0,\Omega_{\lambda},h,\sigma_8)
=(1.0,0.0,0.5,0.56)$ and $(0.3,0.7,0.7,1.0)$, respectively. 
We assume the Harrison-Zel'dovich initial power spectrum.
The sets of the cosmological parameters are chosen so as to reproduce the 
observed cluster abundance (Kitayama \& Suto 1997). And we use the CDM 
transfer function (Bardeen et~al.1986; Sugiyama 1995) with 
$\Omega_bh^2=0.015$ in this section.
Then the CDM power spectrum at present time is written as
\begin{equation}
   P(k)=B k T(k)^2,
\end{equation}
where
\begin{equation}
   T(k)={\ln[1+2.34q]\over 2.34q[1+3.89q+(16.1q)^2+(5.46q)^3+(6.71q)^4]^{1/4}}
\end{equation}
and
\begin{equation}
   q=\biggl({2.726\over 2.7}\biggr)^2
  {k\over \Omega_0h\exp[-\Omega_b-\sqrt{2h}\Omega_b/\Omega_0] 
  h {\rm Mpc}^{-1}},
\end{equation}
where $B$ is a normalization constant.

\subsection{Validity of approximation and implications}

\begin{figure}[t]
\centerline{\epsfxsize=15cm \epsffile{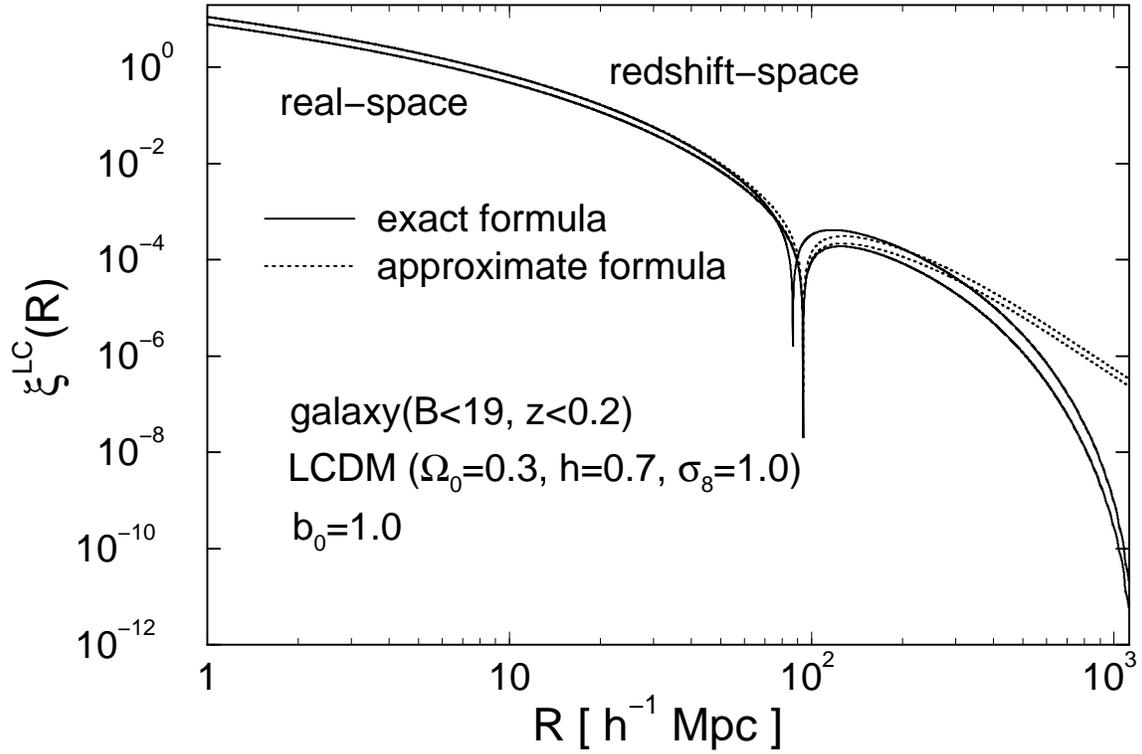}}
\caption{
The two-point correlation functions on a light cone for
galaxy sample in real and redshift spaces. The solid 
line represents the exact formula (29) and the dotted 
line represents the approximate formula (34).
Upper pair curves correspond to the case that
the linear redshift-space distortion is taken into 
account by using formulae  (29) and (34).
Lower pair curves correspond to the case of the real space
by setting $\beta=0$ in (29) and (34).
}
\label{fig1}
\end{figure}

To compute the exact power spectrum on a light-cone, we first perform 
integration of equation ($\ref{eq68.5}$), and obtain the 
two-point correlation function $\xiLC(R)$. Then we
compute the power spectrum by using equation ($\ref{eq13}$). 
As an example to show typical behavior of the two-point correlation 
function, Figure $\ref{fig1}$ plots the galaxy two-point 
correlation function in the LCDM model. Here we adopt the 
case of no bias, i.e., $b_0=1$ in (\ref{bias}).
The solid line, which represents the exact two-point correlation 
function (\ref{eq68.5}), is compared with the dotted line, 
which represents the approximate formula (\ref{xisapprox}). 
The exact two-point correlation function shows a good correspondence
with the approximate formula at small $R$. However the deviation
becomes significant at large $R$ and the exact two-point correlation 
function drops down to zero at the point, where $R$ becomes two 
times of the depth of the survey volume (diameter of the 
survey volume), i.e.,  $2r_{\rm max}(z_{\rm max})$. This is traced back
to our definition of the two-point correlation function. Namely
number density outside the survey region is zero, then a
product of the number density of the two points being 
$R>2r_{\rm max}(z_{\rm max})$ must be zero.

\begin{figure}[t]
\centerline{\epsfxsize=15cm \epsffile{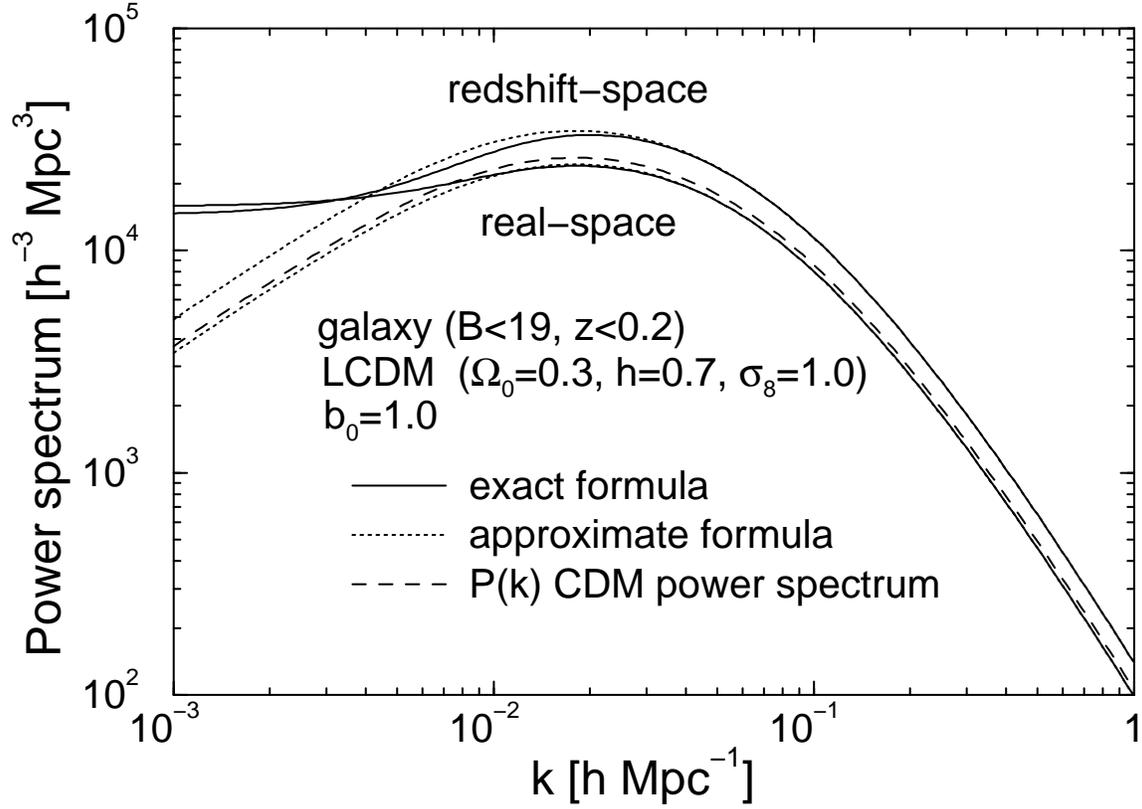}}
\caption{
The power spectra on a light cone for galaxy sample 
in real and redshift spaces. The solid line 
represents the exact power spectrum and the dotted 
line represents the approximate formula (36).
Upper pair curves correspond to the case that
the linear redshift-space distortion is taken into account, 
$P^{\rm LC}_{\rm S}(k)$. 
Lower pair curves correspond to the case in real space, 
$P^{\rm LC}_{\rm R}(k)$.
(Here we refer the 'real space' to the case $\beta(k;\eta)=0$.
in the formulae.) The dashed line corresponds to the CDM power 
spectrum defined on a constant-time hypersurface at present $P(k)$. 
}
\label{fig2}
\end{figure}

Now we discuss the power spectrum. Figure $\ref{fig2}$ plots 
the galaxy power spectra to show characteristic behaviors, 
in which we adopted the LCDM model and the case of no bias.
The solid line, which represents the exact power spectrum, 
is compared with the dotted line, which represents the 
approximate formula (\ref{eq80}).
Comparing the exact and the approximate power spectra, we see 
that the approximation is fairly good for $k\simgt 0.01 h{\rm Mpc^{-1}}$. 
On the larger scales, the exact power spectrum 
approaches a constant value, and the deviation between the exact and
the approximate formulae becomes large. This behavior originates
from the fact that the corresponding wavelength becomes 
larger than depth of the survey volume at the large scales, 
$1/k\simgt r_{\rm max}$, and that we cannot properly evaluate 
the power spectrum because of the finite size effect of the survey volume. 
For comparison, the CDM power spectrum defined 
on a constant-time hypersurface at present is plotted (dashed line). 
Note that in this case the CDM power spectrum is identical to the 
galaxy power spectrum on the constant-time hypersurface because 
we have adopted the case of no bias. 

\begin{figure}[t]
\centerline{\epsfxsize=15cm \epsffile{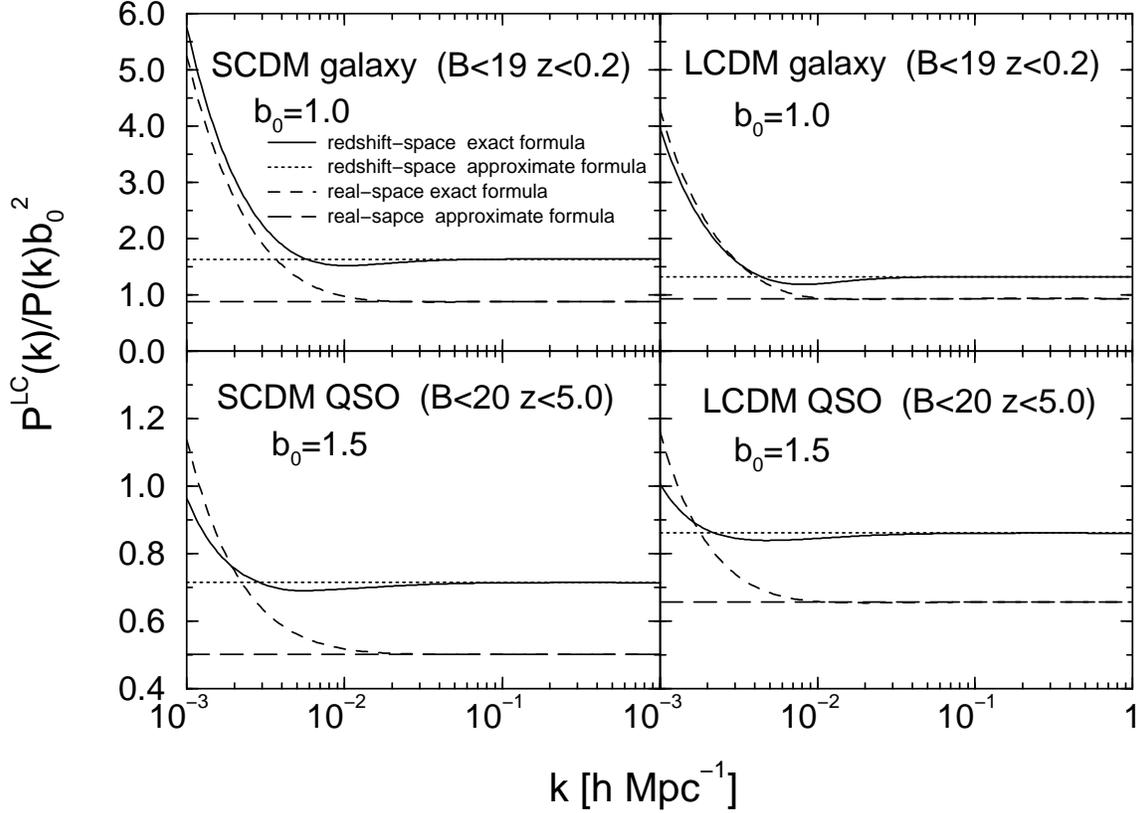}}
\caption{
Real and redshift space power spectra for galaxy and 
quasar samples, divided by power spectra of galaxy 
and quasar on a constant-time hypersurface at present, 
for SCDM and LCDM models. 
The solid line and dotted line represent the exact and approximate
power spectra in redshift space, respectively. 
The short and long dashed lines express 
the exact and the approximate power spectra in real space, respectively.
We adopted the bias model with $b_0=1$ for galaxies, and $b_0=1.5,~p=1$ 
for quasars. 
}
\label{fig3}
\end{figure}

\begin{figure}[t]
\centerline{\epsfxsize=15cm \epsffile{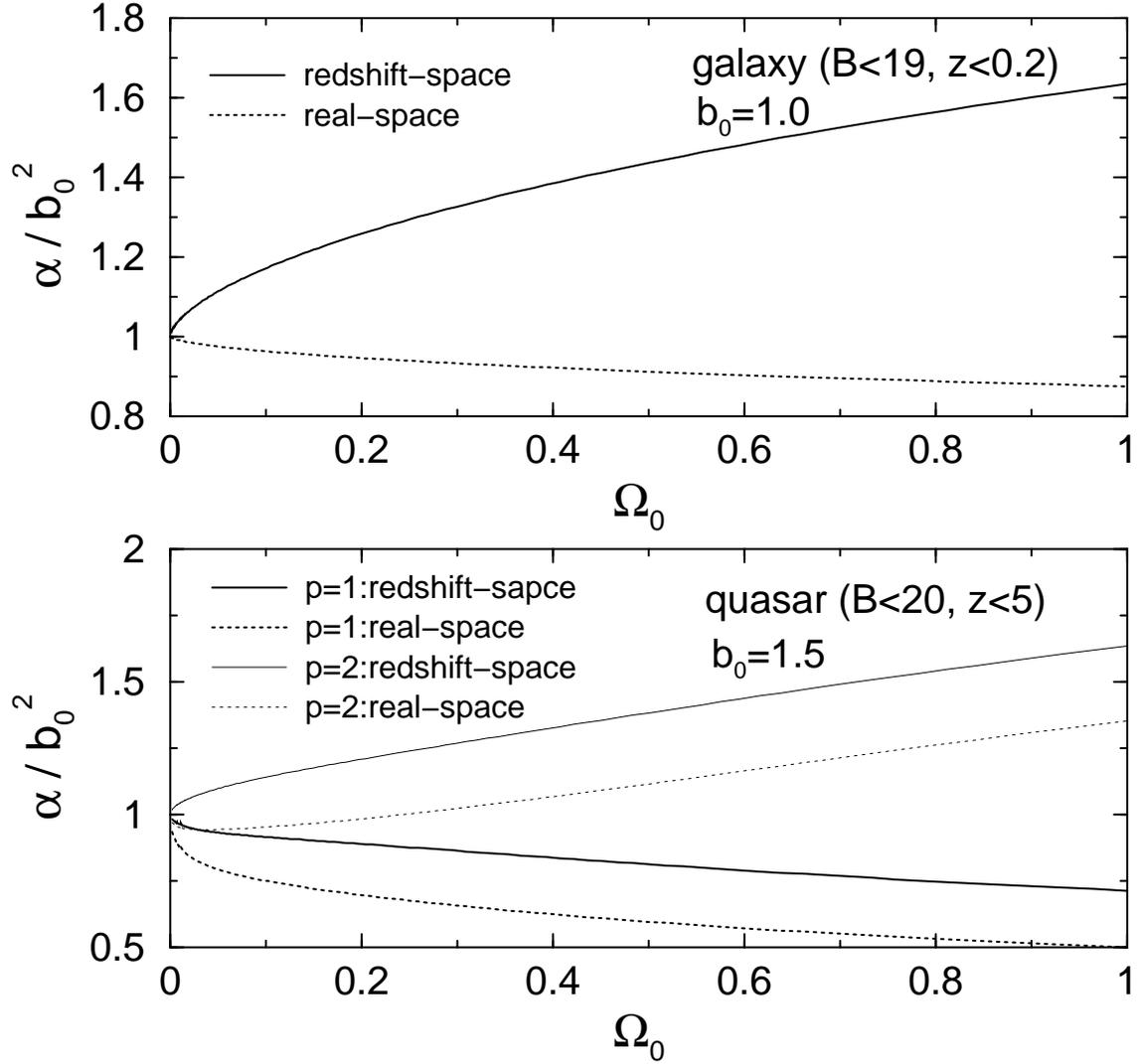}}
\caption{
The factor $\alpha/b_0^2$ for galaxy and quasar as a function
of $\Omega_0$. The solid and the dotted lines represent the 
factor in redshift and real spaces, respectively. 
The case of real space is defined by the case $\beta=0$
in the formulae. 
As for the bias, we adopt the model 
with $b_0=1$ for galaxy, and $b_0=1.5,~{\rm and}~p=1,~2$ for quasar. 
}
\label{fig4}
\end{figure}

To quantify validity of the approximate formula, Figure 
$\ref{fig3}$ plots the power spectra for the galaxy and quasar 
samples, 
which are divided by the power spectrum on the constant-time
hypersurface at present.
The solid line expresses the exact power spectrum and the dotted 
line does the approximate one (\ref{eq80}) in redshift space.
For comparison we show the case in real space, where 
we refer the 'real space' to the case $\beta(k;\eta)=0$ in the formulae.
The short and the long dashed lines express the exact and the approximate
power spectra, respectively, in real space.
It is apparent from Figure $\ref{fig3}$ that the galaxy power spectrum is 
increased by the linear redshift-space distortion which dominantly 
contributes to the galaxy sample. The light-cone effect is not so
effective for the galaxy sample because the sample is shallow.
Nevertheless it can be notable that the light-cone effect decreases
the amplitude of the power spectrum by order of several percent.
On the other hand, the quasar power spectrum is significantly 
decreased because the light-cone effect is effective for such a deep 
observational field. However we should remind that the behavior 
of the power spectrum significantly depends on the time-dependence
of the bias, which determines the amplitude of clustering at high-redshift
(see also Figure {\ref{fig4}}).
%
In each panel, we see that the approximation reproduces the exact 
power spectra fairly well for $k\simgt 0.01 h{\rm Mpc^{-1}}$ for 
the galaxies and quasars samples. 
Thus we conclude that equation ($\ref{eq80}$) provide a good 
approximate formula for the power spectrum on the light cone 
for $k\simgt 0.01 h{\rm Mpc^{-1}}$. 

Here we briefly mention the nonlinear effects on the power spectrum.
While the nonlinear evolution of density field enhances the amplitude, 
the finger-of-God due to the random motion decreases the amplitude
in redshift space. According to the result in Paper ${\III}$, 
the amplitude is decreased by order of several 
$\times 10\%$ relative to its counterpart in linear theory
for both galaxy and quasar samples, depending on the scale $k$ and the 
cosmological model.
This situation is clearly shown in Figure 2 in Paper {\III},
from which we see that the nonlinear effects are effective 
for $k\simgt 0.1h{\rm Mpc}^{-1}$.
Because we have not taken the nonlinear effects into account
throughout the present paper, Figure ${\ref{fig3}}$ in the 
present paper does not show the decrease of the amplitude at 
the small scales. The difference between the Figure ${\ref{fig3}}$
and Figure 2 in Paper {\III} at small $k$ is caused by the finite 
size effect of survey volume.

Finally in this section we discuss dependence of cosmology.
Figure {\ref{fig4}} plots the factor $\alpha/b_0^2$ 
as a function of $\Omega_0$, where $\alpha$
is defined by equation (\ref{defalpha}).
The solid and the dotted lines represent the factor $\alpha/b_0^2$
in redshift space and real space, respectively.
Here we adopt the same model for the galaxy and quasar samples
as in Figure $\ref{fig3}$.
From Figure {\ref{fig4}} it is apparent that the light-cone effect and 
the linear redshift-space distortion becomes more influential as $\Omega_0$ 
increases. This will be traced back to the linear growth rate of density 
perturbations. For the galaxy sample, the increase of amplitude 
due to the redshift-space distortion is more effective than the 
decrease due to the light-cone effect. 
In contrast to this, for the quasar sample, the light-cone effect and 
the time-evolution of the bias become important.

\vspace{1cm}

\section{COSMOLOGICAL REDSHIFT-SPACE DISTORTION}

\def\n0LC{{{\tilde n}^{\LC}}}
\def\bfs{{\bf s}}
\def\bfr{{\bf r}}
\def\bfR{{\bf R}}
\def\Omegar{{\Omega}}
\def\Omegas{{\bar\Omega}}
\def\ar{{a}}
\def\as{{\bar a}}
\def\chir{{\chi}}
\def\chis{{\bar\chi}}
\def\etar{{\eta}}
\def\etas{{\bar\eta}}
The crucial point of our investigation in the previous section
is the assumption that we know the exactly correct cosmological 
model of the universe. 
Because the redshift $z$ and the direction $\vect{\gamma}$ are the
observable quantities of objects in the redshift survey, 
then we must assume a cosmological model to obtain a three-dimensional 
map on which the cosmological objects are plotted. 
Since the cosmological parameters of our universe have not been 
established completely, then we should be careful for this uncertainty. 

It is known that an apparent shape of distribution of 
cosmological objects is distorted by the cosmological
redshift-space distortion.
The cosmological redshift-space distortion has been 
discussed as a tool for the cosmological model (Alcock \& 
Paczynski 1979; Ryden 1995; Ballinger, Peacock, 
\& Heavens 1996; Matsubara \& Suto 1996; Suto, et~al. 1999; 
Magira, Jing, \& Suto 1999).
An incorrect assumption of the cosmological model affects 
proper estimation of the two-point correlation 
function and the power spectrum. In this section we focus on 
this point and we investigate how the error of the assumption of 
cosmological model 
affects the estimation of the power spectrum on a light cone. 

\subsection{Basic Formulation}
In this section, we assume that the universe is the Friedmann-Lemaitre 
universe with a cosmological constant and that the 'correct' matter 
density parameter is $\Omegar_0$. We also assume that data processing 
is performed by assuming the Friedmann-Lemaitre universe 
with an 'incorrect' matter density parameter $\Omegas_0$.
In this section we neglect the effect of the linear redshift-space
distortion for simplicity. In this case, the real space is the universe 
with the 'correct' density parameter $\Omega_0$, and is described by 
the same relations in the previous section. And we use the same notations 
from equation (\ref{eq14}) to (\ref{eq21}) to describe the real space.
On the other hand, the redshift space is the universe with the 
'incorrect' density parameter $\Omegas_0$. To describe the 
redshift space, we write the line element
\begin{equation}
  ds^2 = \as^2(\etas) 
  \left[-d\etas^2+d\chis^2+\chis^2 d\Omega_{(2)}^2 \right],
\label{metrics}
\end{equation}
where $\as(\etas)$ is the scale factor normalized to be unity at 
present, i.e., $\as(\etas_0)=1$, 
$\etas$ is the conformal time, $\chis$ is the radial coordinate,
and and $d\Omega_{(2)}^2$ denotes the line element of the unit 
two-sphere.
Evolution of the scale factor $\as$ is specified by 
the similar equation as (\ref{eq15}), 
\begin{equation}
  \biggl({1\over \as}{{d \as}\over d\etas}\biggr)^2={\bar H}_0^2\biggl(
  {\Omegas_0\over \as}+\as^2\Omegas_{\lambda}\biggr),
\label{eqs15}
\end{equation}
where $\Omegas_{\lambda}=1-\Omegas_{0}$, and ${\bar H}_0
=100{\bar h}{\rm km/s/Mpc}$ is the Hubble parameter in redshift space. 
As in subsection 3.1, we assume
that an observer is located at the origin of the coordinates 
$(\etas=\etas_0,\chis=0)$, and the light-cone hypersurface of 
the observer satisfies the relation $\etas=\etas_0-\chis$. 
Then we introduce the radial coordinate $s$ to denote the metric 
of the three dimensional redshift-space on the light-cone hypersurface as
\begin{equation}
  d\bar s^2_{\rm LC}=ds^2+s^2 d\Omega_{(2)}^2.
\label{ds}
\end{equation}

The relation between $z$ and $s$ is specified by $\as(\etas_0-s)=1/(1+z)$. 
When a set of data 
$(z,\vect{\gamma})$ for cosmological objects was obtained,
we compute the number density field $F(s,\vect{\gamma})$ 
defined by equation (\ref{eq1}) with the number density field 
$n^\RS(s,\vect{\gamma})$ and that for a synthetic catalog 
$n^\RS_{\rm syn}(s,\vect{\gamma})$. 
And a map is constructed in redshift space (\ref{ds}).
Then we can compute the two-point correlation function and the power 
spectrum, according as the definitions (\ref{eq9}) and (\ref{eq7}), 
respectively.
We specify relations between the redshift space and the real 
space, as is done in subsection 3.1. The relation for the number density 
field is specified by the same equations as (\ref{eq23}) and (\ref{eq28}).
The relation between $s$ and $r$ is specified by
\begin{equation}
  a(\eta_0-r)=\as(\etas_0-s)={1\over 1+z},
\label{aas}
\end{equation}
where $a(\eta)$ and $\as(\etas)$ are the solutions of
equations (\ref{eq15}) and (\ref{eqs15}), respectively.

Now we calculate the two-point correlation function.
With the use of equations, (\ref{eq23}), (\ref{eq28}), and (\ref{eq19}),
we find that (\ref{eq9}) reduces to
\begin{eqnarray}
  \xiLC(R)=
 {
   {\displaystyle
   \int {d\Omega_{\hat{\vect{R}}}\over 4\pi}
  \int d^3\vect{r}_1\int d^3\vect{r}_2
  {\tilde n}^{\LC}(r_1){\tilde n}^{\LC}(r_2)
  \Bigl< \Delta^{\LC}(r_1,\vect{\gamma}_1)
         \Delta^{\LC}(r_2,\vect{\gamma}_2)\Bigr>
      \delta^{(3)}(\vect{s}_1-\vect{s}_2-\vect{R})}
\over
    {\displaystyle
   \int d^3 \vect{r} \biggl({dr\over ds}\biggr) \biggl({r\over s}\biggr)^2
   {\tilde n}^{\LC}(r)^2}
 }~,
\end{eqnarray}
where we used ${\tilde n}^\RS(s) s^2 ds={\tilde n}^{\LC}(r)r^2 dr$
to write the denominator.

After straightforward calculations similar to those in Appendix, 
we find that the above expression reduces to
\begin{eqnarray}
  &&\xiLC(R)=\biggl[4\pi\int d r r^2 \biggl({dr\over ds}\biggr) 
  \biggl({r\over s}\biggr)^2 {\tilde n}^{\LC}(r)^2\biggr]^{-1}
\nonumber
\\
  &&\hspace{1cm}\times {1\over \pi R}
  \int\int_{\cal M} dr_1 dr_2  
  \prod_{j=1}^2
  \biggl({r_j^2\over s_j}{\tilde n}^{\LC}(\eta_0-r_j) D_1(\eta_0-r_j)\biggr)
\nonumber
\\
  &&\hspace{1cm}\times \int dk k^2  P(k) b(k;\eta_0-r_1) b(k;\eta_0-r_2)
  j_0\Biggl(k\sqrt{r_1^2+r_2^2-r_1r_2{s_1^2+s_2^2-R^2 \over s_1s_2}}\Biggr)
  ~,
\label{C66}
\end{eqnarray}
where $\cal M$ denotes the region $|s_1-s_2|\leq R\leq 
s_1+s_2$, and $s_1$ and $s_2$ are understood as functions of 
$r_1$ and $r_2$, respectively, which are specified by equation 
(\ref{aas}). Thus the formula (\ref{C66}) is the 
exact two-point correlation function which takes 
the cosmological redshift-space distortion  into account.
Note that (\ref{C66}) reduces to the two-point correlation function
in real space $\xi^{\rm LC}_{\rm R}(R)$, which is explicitly defined 
by (\ref{eq68.5}) with setting $\beta(k;\eta)=0$,  when the 'correct' 
cosmological model is chosen as the redshift space, i.e., 
$\Omega_0=\Omegas_0$ and $h={\bar h}$. 

\subsection{Distant observer approximation}
We consider an approximate expression of (\ref{C66}) in the simlar
way to subsection 3.2.
We introduce the variables $y=s_2-s_1$ and $x=r_2-r_1$, and set that 
$y\ll s_1, s_2$ and $x\ll r_1, r_2$. In this case the argument 
of the spherical bessel function in (\ref{C66}) is approximated as
\begin{equation}
  k\sqrt{r_1^2+r_2^2-r_1r_2{s_1^2+s_2^2-R^2 \over s_1s_2}}
  \simeq
  k\sqrt{{{r_1^2\over s_1^2}R^2+\biggl[\biggl({dr_1\over d s_1}\biggr)^2
  -{r_1^2\over s_1^2}\biggr]y^2}},
\end{equation}
where we used
\begin{equation}
  {x\over y}={r_2-r_1\over s_2-s_1}\simeq {dr_1\over ds_1}.
\end{equation}
Taking $s_1$ and $y$ as the integration variables
instead of $r_1$ and $r_2$ in (\ref{C66}), and we 
use the approximation,
\begin{equation}
  \int\int_{\cal M} ds_1 ds_2  \simeq \int ds_1\int_{-R}^R dy.
\end{equation}
Introducing the new variable $\mu$ instead of $y$ by $\mu=y/R$, 
then (\ref{C66}) is approximated as
\begin{eqnarray}
  &&\xiLC(R)\simeq\biggl[4\pi\int d s s^2 \biggl({dr\over ds}\biggr)^2 
  \biggl({r\over s}\biggr)^4 {\tilde n}^{\LC}(r)^2\biggr]^{-1}
\nonumber
\\
  &&\hspace{1cm}\times {1\over \pi}
  \int ds_1\int_{-1}^1 d\mu  \biggl({dr_1\over ds_1}\biggr)^2
  \biggl({r_1^2\over s_1}\biggr)^2
  \Bigl[{\tilde n}^{\LC}(\eta_0-r_1) D_1(\eta_0-r_1)\Bigr]^2
\nonumber
\\
  &&\hspace{1cm}\times \int dk k^2  P(k) b(k;\eta_0-r_1)^2
  j_0\Bigl(kR  \Theta \Bigr),
\label{D2}
\end{eqnarray}
where we defined 
\begin{equation}
  \Theta=\sqrt{{{r_1^2\over s_1^2}+\biggl[\biggl({dr_1\over d s_1}\biggr)^2
  -{r_1^2\over s_1^2}\biggr]\mu^2}}.
\end{equation}

Substituting equation (\ref{D2}) into (\ref{eq13}), one obtains 
\begin{eqnarray}
  &&\PLC(k)\simeq\biggl[\int d s s^2 \biggl({dr\over ds}\biggr)^2 
  \biggl({r^2\over s^2}\biggr)^2 {\tilde n}^{\LC}(r)^2\biggr]^{-1}
\nonumber
\\
  &&\hspace{1cm}\times 
  \int ds_1s_1^2\biggl({dr_1\over ds_1}\biggr)^2
  \biggl({r_1^2\over s_1^2}\biggr)^2
  \Bigl[{\tilde n}^{\LC}(\eta_0-r_1) D_1(\eta_0-r_1)\Bigr]^2
\nonumber
\\
  &&\hspace{1cm}\times \int_{0}^1 d\mu 
  P\biggl({k\over \Theta}\biggr)
 { b\Bigl(k/\Theta;\eta_0-r_1\Bigr)^2\over \Theta^3}~.
\label{D3}
\end{eqnarray}
It might be more useful to rewrite this expression in terms of the 
mean number density per unit redshift and per unit solid angle 
$dN(z)/dz$, which is defined by 
$dN(z)={\tilde n}^\RS(s)s^2ds={\tilde n}^{\LC}(r)r^2 dr$,
\begin{eqnarray}
  \PLC(k)\simeq
 {
   {\displaystyle
  \int dz \biggl({dz\over ds}\biggr){1\over s^2}
  \biggl({dN\over dz}\biggr)^2 D_1[z]^2
  \int_{0}^1 d\mu P\biggl({k\over \Theta}\biggr){
  b\Bigl[k/\Theta;z\Bigr]^2
  \over \Theta^3}
  }
\over
    {\displaystyle
  \int dz \biggl({dz\over ds}\biggr){1\over s^2}
  \biggl({dN\over dz}\biggr)^2
  }
 }~,
\label{usefulD3}
\end{eqnarray}
where $s$ is regarded as a function of $z$ specified by 
the relation $\as(\etas_0-s)=1/(1+z)$.

\subsection{Validity of approximation and implications}
We show how the cosmological redshift-space distortion distorts
the power spectrum on a light cone. In this section we adopt
the same galaxy and quasar samples as in subsection 3.4.
We take $\Omega_0=0.3$ as the density parameter of  
the real space (the 'correct' universe). 
And we adopt $\Omegas_0=1.0$ as that of the redshift space.
In this section, we assume that $h={\bar h}=0.7$ 
for the real and the redshift spaces for simplicity. 
\footnote{ The difference of the Hubble parameter can 
cause an additional difference in the power spectra
which is not essential to the cosmological redshift-space 
distortion. Therefore we here choose $h={\bar h}$ for definiteness.}

\begin{figure}[t]
\centerline{\epsfxsize=10cm \epsffile{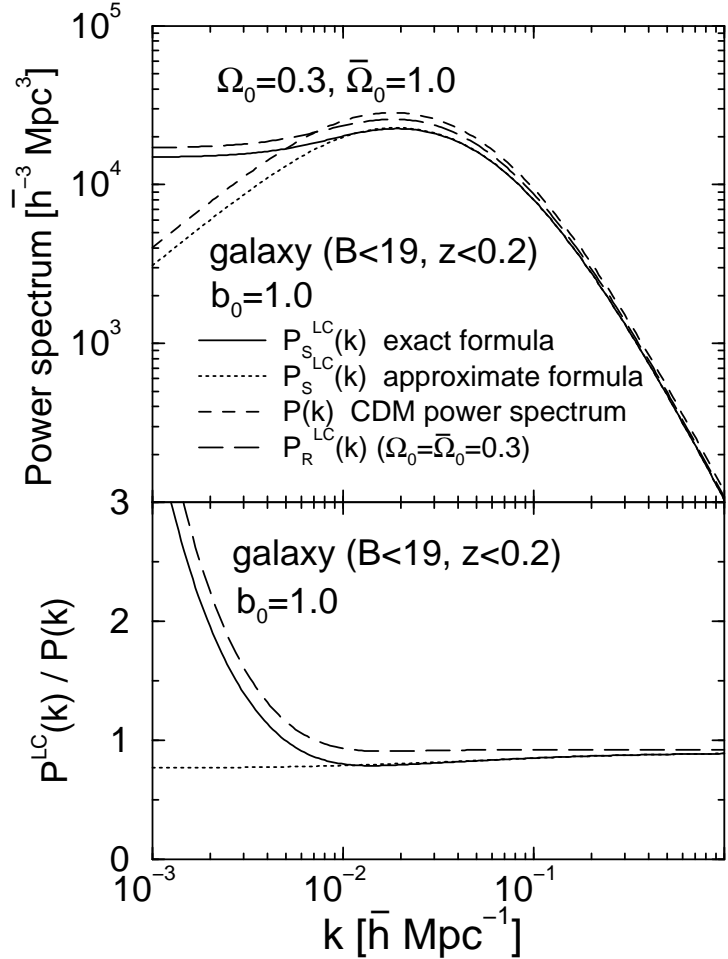}}
\caption{
Galaxy power spectrum on a light cone which 
incorporates the cosmological redshift-space distortion.
Here we adopted the galaxy sample in subsection 3.4, 
and $(\Omega_0=0.3,h=0.7)$, and $({\bar \Omega}_0=1.0,~{\bar h}=0.7)$. 
The case of no bias is considered.
The solid and dotted line represent the exact and the approximate 
power spectra, respectively. The short dashed line plots the 
power spectrum on a constant-time hypersurface and the 
long dashed line labeled by $P^{\rm LC}_{\rm R}(k)$ does the power spectrum 
assuming the correct cosmological model in redshift space, 
i.e.,  $(\Omega_0={\Omegas}_0=0.3)$. 
Thus the long dashed line merely expresses the decrease of
the power spectrum due to light-cone effect. 
The upper panel shows the power spectra, and the lower panel 
shows the power spectra divided by the power spectrum on the
constant-time hypersurface at present, $P(k)$.
}
\label{fig5}
\end{figure}

\begin{figure}[t]
\centerline{\epsfxsize=15cm \epsffile{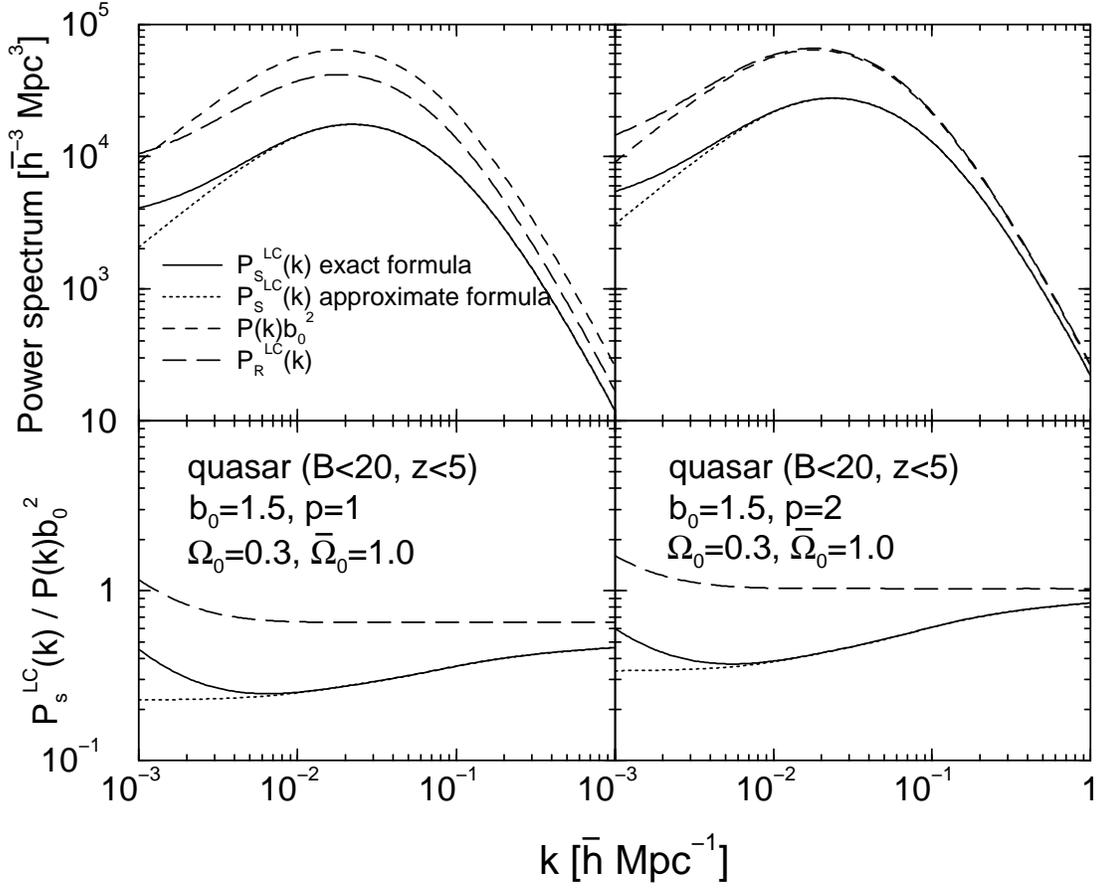}}
\caption{
Quasar power spectra on a light cone which take the 
cosmological redshift-space distortion into account.
We adopted the quasar sample in subsection 3.4, 
and $\Omega_0=0.3$,~$h=0.7$, and ${\bar \Omega}_0=1.0$,~${\bar h}=0.7$.
Here the bias model with $b_0=1.5,~{\rm and}~p=1,~2$ is considered.
The meanings of the lines are the same as Figure 5.
}
\label{fig6}
\end{figure}

The exact power spectrum which takes the cosmological redshift-space 
distortion into account is computed from equation (\ref{eq13}) by 
integrating the exact two-point correlation function (\ref{C66}). 
The approximate formula is given by equation (\ref{D3}) or (\ref{usefulD3}). 
Figure $\ref{fig5}$ plots the galaxy power spectra.
Figure $\ref{fig6}$ shows the case of the quasars.
The solid and the dotted lines represent the exact and the approximate 
power spectra, respectively.
It is apparent from the figures 
that the approximate formula shows the good correspondence
with the exact formula for $k\simgt0.01 h{\rm Mpc}^{-1}$. 
The exact power spectrum approaches a constant value 
at the large scale, and the deviation of the exact formula
from the approximate formula becomes large for 
$k\simlt0.01~h{\rm Mpc}^{-1}$.
This is caused by the finite size effect of the survey volume, 
as mentioned in subsection 3.5.

In Figures \ref{fig5} and \ref{fig6}, the short dashed line shows the 
power spectrum defined on the constant-time hypersurface at present 
$P(k)b_0^2$. In lower panels, the power spectrum,
divided by the power spectrum at present, $\PLC(k)/P(k)b_0^2$, is 
plotted. Although we here adopted the scale-independent bias model,
the distortion of the power spectrum depends on the scale $k$.
The location of the peak of the power spectrum is moved, due to 
the redshift-space distortion.
Thus the cosmological redshift-space distortion distorts the power spectrum 
by shifting the wavenumber $k$, which is expected from 
the expression (\ref{D3}) or (\ref{usefulD3}). 
This feature is in marked contrast to 
the case of the linear redshift-space distortion (see equation 
[\ref{eq80}] and Figure \ref{fig3}).

The long dashed line plots the case $\Omega_0={\bar \Omega}_0=0.3$,
which is labeled by $P^{\rm LC}_{\rm R}(k)$. That is, the 
long dashed line shows the case when the cosmological 
redshift-space distortion is not effective, and it merely 
shows a contribution from the light-cone effect. 
Comparing the solid line and the long dashed line
in the lower panels, we conclude that the cosmological 
redshift-space distortion is an important effect not only 
for the quasar sample but also for the galaxy sample in
estimating the power spectrum. 
The cosmological redshift-space distortion affects the
power spectrum at the $10$ percent level for the galaxy sample.
For the quasar samples, the effect becomes more significant. 
Moreover we should note that the behavior significantly depends on 
the time-evolution of the bias model.

\section{CONCLUSIONS}

In the present paper, we investigated the power spectrum of 
cosmological objects on a light cone focusing on the redshift-space
distortions in linear theory of density perturbations.
We developed theoretical formulations to compute the power 
spectrum which properly takes into account the light-cone effect 
and the redshift-space distortion effects. 
The linear and the cosmological redshift-space distortions
were considered in section 3 and 4, respectively.
The formulae for the power spectra were derived in the rigorous 
manner starting from the first principle corresponding to the Fourier 
analysis. Using the distant observer approximation, we derived 
the approximate formulae (eqs.[\ref{eq80}] and [\ref{D3}]),  
and examined the validity and the limitations. It was shown that 
the approximate expressions are useful to describe the effects 
properly.

Applying our formulae to galaxy and quasar samples which roughly 
match the SDSS survey, we showed that the effects considered in the 
present paper become important for on-going wide and 
deep surveys like the SDSS and 2dF surveys.
Let us summarize the results and the implications 
obtained in this paper. For the galaxy 
samples, the light-cone effect decreases the amplitude of
the power spectrum several percentage points; 
the cosmological redshift-space distortion can distort the
power spectrum by order of ten percent; and the 
linear redshift-space distortion increases the amplitude 
by order of several $\times10$ percent, depending 
on the cosmological parameters. The power spectrum is rather 
insensitive to the time-evolution of the bias model 
for the shallow galaxy samples (Paper $\III$).
For the quasar samples, the linear redshift-space distortion 
can be a minor effect, and the light-cone effect and the 
cosmological redshift-space distortion are the influential 
effects. Of course the power spectrum significantly depends 
on the cosmological model and the bias model.

In the present paper we have shown the validity of using the 
approximate formulas at the small scales. Our numerical calcuations 
have shown that the approximate formulas become invalid at the large 
scales for $k\simlt 0.01 h{\rm Mpc}^{-1}$. Let us briefly
discuss the physical meaning of the discrepancy between the
exact formula and the approximate formula. The terminology, exact,
means that the formulas are derived from the definitions of 
the statistical estimator for the two-point correlation function 
and the power spectrum, which corresponds to a conventional 
data processing. Then the exact formulas
indicate how the finite size effect of the survey volume
affects the two-point correlation function and the power spectrum.
For the two-point correlation function the finite size effect
decreases the amplitude of the correlation function  at the 
large scales. For example the amplitude becomes zero for 
$R\geq 2r_{\rm max}$, where $2r_{\rm max}$ is the diameter
of the survey volume (see Figure 1). For the large separation 
(for large $R$), the sign of the two-point correlation function 
is negative, which implies the anti-correlation at the large scales. 
Thus the finite size effect decreases the anti-correlation at the
large scales. We suppose that this causes the large amplification 
of the power spectra at the large scales (at small $k$) 
relative to its counterpart in linear theory. 

Because we worked in linear theory in the present paper, 
the behaviors on small scales (at large $k$) are not
correctly described, where the nonlinear effect 
causes an additional distortion of the power spectrum.
The distortion of the power spectrum was discussed by 
extending the formula to incorporate the nonlinear effect 
in Paper {\III}, though the exact derivation was invalidated 
by the inclusion of the nonlinear effect in a strict sense. 
According to the result, the nonlinear effect causes the 
additional distortion of the power spectrum
by order of several $\times10$ percent at
$0.1 h{\rm Mpc}^{-1}\simlt k \simlt 1 h{\rm Mpc}^{-1}$,
depending on the cosmological model. 
In the present paper, the linear redshift-space distortion and the
cosmological redshift-space distortion are discussed separately,
however, investigations combining these effects are needed
to present realistic theoretical predictions.
(Suto, Magira, \& Yamamoto 1999).  
Moreover we note that the bias model is the crucial problem to 
present precise theoretical predictions for the power spectrum 
(Moscardini, et~al. 1998; Dekel \& Lahav 1999; Taruya, Koyama, 
\& Soda 1999; Tegmark \& Peebles 1998), though we here adopted 
a simple model.

\bigskip

\bigskip
\bigskip
\begin{center}
{\bf ACKNOWLEDGMENTS}
\end{center}

We are grateful to Yasufumi Kojima for useful discussions and comments. 
We thank Yasushi Suto for his useful discussions, instructions, and 
for providing numerical routines to compute quasar selection functions.
We also thank Silvio Perez for carefully reading the manuscript.
Numerical computations were carried out in part on INSAM of Hiroshima 
University.
This research was supported by the Inamori Foundation and in part by the 
Grants-in-Aid program (11640280) by the Ministry of Education, Science, 
Sports and Culture of Japan.

\newpage
\bigskip
\bigskip
\begin{appendix}
\begin{center}
{\bf APPENDIX}
\end{center}

\section{Brief summary of the calculation $\xiLC(R)$}
We here briefly outline the calculation of equation (\ref{eq68.5}).
We use the almost same mathematical notations as those in Paper \II, 
in which similar calculations have been given.  The calcuations
in Paper \II~ will supplement arguments in this Appendix.
We work within a framework of linear theory of density perturbations
based on the CDM cosmological model. We expand the CDM density 
contrast $\delta_c$ and the velocity field $v_c^i$ in terms of 
the scalar harmonics as follows (e.g., Kodama \& Sasaki 1984)
\begin{eqnarray}
  &&\delta_c(\eta,\chi,\vect{\gamma})
  =\int_0^\infty dk \sum_{l,m} \delta^{\rm (c)}_{klm}(\eta)
    {\cal Y}_{klm}(\chi,\vect{\gamma}),
\label{eq38}
\\
  &&v^i_c(\eta,\chi,\vect{\gamma})=\int_0^\infty dk \sum_{l,m} 
  {\dot\delta^{\rm (c)}_{klm}(\eta) \over k^2}
  {\cal Y}_{klm}(\chi,\vect{\gamma}){}^{|i},
\label{eq49}
\end{eqnarray}
where ${\cal Y}_{klm}(\chi,\vect{\gamma})$ is the normalized scalar harmonics
\begin{eqnarray}
  &&{\cal Y}_{klm}(\chi,\vect{\gamma})= 
X_{k}^{l}(\chi) Y_{lm}(\Omega_{\vect{\gamma}}),
\label{eq40}
\end{eqnarray}
with 
\begin{equation}
   X_{k}^{l}(\chi)=\sqrt{{2\over \pi}} k j_l(k\chi),
\label{eq41}
\end{equation}
and $Y_{lm}(\Omega_{\vect{\gamma}})$ and $j_l(x)$ are the spherical 
harmonics and the spherical Bessel function, respectively, and 
${\cal Y}_{klm}(\chi,\vect{\gamma}){}^{|i}$ denotes the 
covariant derivative of ${\cal Y}_{klm}(\chi,\vect{\gamma})$ 
on the three-dimensional space. The Fourier coefficient 
$\delta^{\rm (c)}_{klm}(\eta)$ satisfies 
\begin{equation}
  \ddot\delta^{\rm (c)}_{klm}+{\dot a\over a}\dot\delta^{\rm (c)}_{klm}
  -{3\over2}{\Omega_0H_0^2\over a}\delta^{\rm (c)}_{klm}=0.
\label{eq46}
\end{equation}
In the Friedmann-Lemaitre universe, the growing mode solution is well
known:
\begin{equation}
  \delta^{\rm (c)}_{klm}(\eta)=\delta^{\rm (c)}_{klm}(\eta_0)D_1(a),
\label{eq47}
\end{equation}
with
\begin{equation}
  D_1(a)={A}\sqrt{{\Omega_0\over a^3}+1-\Omega_0}
  \int_0^a da'\biggl({a'\over \Omega_0+a'^3(1-\Omega_0)}\biggr)^{3/2}.
\label{eq48}
\end{equation}
Here ${A}$ is a constant to be determined so that $D_1$ 
is unity at present.

We assume that the number density contrast of objects connects with
the CDM density contrast by the scale-dependent bias factor 
$b(k;\eta)$ as,
\begin{equation}
  \Delta_{klm}(\eta)=b(k;\eta)\delta^{(c)}_{klm}(\eta).
\label{eq50}
\end{equation}
Furthermore we assume that the velocity field of the objects,
$\vect{v}(\eta,\chi,\vect{\gamma})$, is the same as that of CDM
$\vect{v}_c(\eta,\chi,\vect{\gamma})$. In this case the apparent 
shift of comoving distance of an object, $\delta r(r,\vect{\gamma})$, 
is related to the velocity as
\begin{equation}
\delta r(r,\vect{\gamma})={
  a(\eta)^{1\over2}\over H_0\sqrt{\Omega_0+\Omega_{\lambda}a(\eta)^3}}
\vect{\gamma}\cdot\vect{v}_c(\eta,\chi,\vect{\gamma})
 \Big\vert_{\eta\rightarrow \eta_0-r, \chi\rightarrow r}~.
\label{eq25}
\end{equation}
%
With these assumptions we write
\begin{eqnarray}
  &&\Delta^{\LC}(r,\vect{\gamma})=\int^\infty_0 dk \sum_{lm}
  \delta^{(\rm c)}_{klm}(\eta_0)b(k;\eta_0-r)D_1(\eta_0-r)
  {\cal Y}_{klm}(r,\vect{\gamma}),
\label{eq51}
\\
  &&\delta r(r,\vect{\gamma})=
  \int^{\infty}_0 dk \sum_{lm}\delta^{(\rm c)}_{klm}(\eta_0)
  f(\eta_0-r)D_1(\eta_0-r)k^{-2}{\cal Y}_{klm}(r,\vect{\gamma})^{|r},
\label{eq52}
\end{eqnarray}
where
\begin{equation}
 f(\eta)={d\ln D_1(\eta)\over d \ln a(\eta)}~.
\label{eq53}
\end{equation}

Substituting equations (\ref{eq51}) and (\ref{eq52}) into (\ref{eq34}),
we obtain
\begin{eqnarray}
\xiLC(R)&=&\biggl[\int d^3\vect{r}{\tilde n}^{\LC}(r)^2\biggr]^{-1}
\int{d\Omega_{\hat{\vect{R}}}\over 4\pi}
\int dr_1r_1^2d\Omega_{\vect{\gamma}_1}  
\int dr_2r_2^2d\Omega_{\vect{\gamma}_2}
\nonumber \\
&&\times {\tilde n}^{\LC}(r_1){\tilde n}^{\LC}(r_2)D_1(\eta_0-r_1)D_1(\eta_0-r_2)
\nonumber \\
&&\times\int^{\infty}_0dk_1\sum_{l_1m_1}\int^{\infty}_0dk_2\sum_{l_2m_2}
\biggl<\delta^{({\rm c})}_{k_1l_1m_1}(\eta_0)
\delta^{({\rm c})*}_{k_2l_2m_2}(\eta_0)\biggr>
Y_{l_1m_1}(\Omega_{\vect{\gamma}_1})
Y_{l_2m_2}^*(\Omega_{\vect{\gamma}_2})
\nonumber \\
&&\times\prod_{i=1}^2\biggl\{\biggl
  [b(k_i;\eta_0-r_i)-k_i^{-2}{\cal D}_{r_i}\biggr]X^{l_i}_{k_i}(r_i)\biggr\}
  \delta^{(3)}(\vect{r}_1-\vect{r}_2-\vect{R}),
\label{eq54}
\end{eqnarray}
where
\begin{equation}
{\cal D}_r=f(\eta_0-r){d\over dr}\ln \biggl[r^2{\tilde n}^{\LC}(r)f(\eta_0-r)
D_1(\eta_0-r)\biggr]{\partial\over\partial r}+f(\eta_0-r)
{\partial^2\over\partial r^2}.
\label{eq55}
\end{equation}
Using the relation
\begin{equation}
e^{i\vect{k}\cdot\vect{r}}=4\pi\sum_l\sum^l_{m=-l}
(-i)^ij_l(k|\vect{r}|)Y_{lm}(\Omega_{\hat{\vect{k}}})
Y^{*}_{lm}(\Omega_{\hat{\vect{r}}}),
\label{eq56}
\end{equation}
equation ($\ref{eq54}$) is written as
\begin{eqnarray}
\xiLC(R)&=&\biggl[\int d^3 \vect{r}{\tilde n}^{\LC}(r)^2\biggr]^{-1}
\int{d\Omega_{\hat{\vect{R}}}\over 4\pi}
\int dr_1r_1^2d\Omega_{\vect{\gamma}_1}  
\int dr_2r_2^2d\Omega_{\vect{\gamma}_2}
\nonumber \\
&&\times {\tilde n}^{\LC}(r_1){\tilde n}^{\LC}(r_2)D_1(\eta_0-r_1)D_1(\eta_0-r_2)
\nonumber \\
&&\times\int^{\infty}_0dk_1\sum_{l_1m_1}\int^{\infty}_0dk_2\sum_{l_2m_2}
\biggl<\delta^{({\rm c})}_{k_1l_1m_1}(\eta_0)
\delta^{({\rm c})*}_{k_2l_2m_2}(\eta_0)\biggr>
Y_{l_1m_1}(\Omega_{\vect{\gamma}_1})
Y_{l_2m_2}^*(\Omega_{\vect{\gamma}_2})
\nonumber \\
&&\times {2\over\pi}\prod^2_{i=1}\biggl\{\biggl[b(k_i;\eta_0-r_i)
-k_i^2{\cal D}_{r_i}\biggr]k_ij_{l_i}\biggr\}
\nonumber \\
&& \times {1\over (2\pi)^3}\int d^3{\vect{k}}
4\pi\sum_{L_1M_1}(-i)^{L_1}j_{L_1}(kr_1)
Y_{L_1M_1}(\Omega_{\hat{\vect{k}}})
Y_{L_1M_1}^*(\Omega_{\vect{\gamma}_1})
\nonumber \\
&& \times 4\pi\sum_{L_2M_2}(i)^{L_2}j_{L_2}(kr_2)
Y_{L_2M_2}^*(\Omega_{\hat{\vect{k}}})
Y_{L_2M_2}(\Omega_{\vect{\gamma}_2})
\nonumber \\
&& \times 4\pi\sum_{L_3M_3}(i)^{L_3}j_{L_3}(kR)
Y_{L_3M_3}^*(\Omega_{\hat{\vect{k}}})
Y_{L_3M_3}(\Omega_{\hat{\vect{R}}})~,
\label{eq57}
\end{eqnarray}
where $k=\vert\vect{k}\vert$ and $\hat{\vect{k}}=\vect{k}/k$. 
Integration over $\Omega_{\vect{\gamma}_1}$, $\Omega_{\vect{\gamma}_2}$,
$\Omega_{\hat{\vect{R}}}$, and $\Omega_{\hat{\vect{k}}}$ yields
\begin{eqnarray} 
\xiLC(R)&=&\biggl[\int d^3{\bf r}{\tilde n}^{\LC}(r)^2\biggr]^{-1}
\int dr_1r_1^2 \int dr_2r_2^2
{\tilde n}^{\LC}(r_1){\tilde n}^{\LC}(r_2)D_1(\eta_0-r_1)D_1(\eta_0-r_2)
\nonumber \\
&&\times\int^{\infty}_0dk_1\sum_{l_1m_1}\int^{\infty}_0dk_2\sum_{l_2m_2}
\biggl<\delta^{({\rm c})}_{k_1l_1m_1}(\eta_0)
\delta^{({\rm c})*}_{k_2l_2m_2}(\eta_0)\biggr>
\nonumber \\
&&\times {2\over \pi} \prod^2_{i=1}\biggl\{\biggl[b(k_i;\eta_0-r_i)
-k_i^{-2}{\cal D}_{r_i}\biggr]k_ij_{l_i}(k_ir_i)\biggr\}
\nonumber \\
&& \times {(4\pi)^2 \over (2\pi)^3}
\int d k k^2 j_{l_1}(kr_1)j_{l_2}(kr_2)j_0(kR)
\delta_{l_1,l_2}\delta_{m_1,m_2}~.
\label{eq58}
\end{eqnarray}
Because the Gaussian random fluctuations satisfy
\begin{equation}
\biggl<\delta^{({\rm c})}_{k_1l_1m_1}(\eta_0)
\delta^{({\rm c})*}_{k_2l_2m_2}(\eta_0)\biggr>
=\delta(k_1-k_2)\delta_{l_1l_2}\delta_{m_1m_2}P(k_1),
\label{eq59}
\end{equation}
where $P(k)$ is the CDM power spectrum at present, we obtain
\begin{eqnarray}
\xiLC(R)&=&\biggl[\int d^3\vect{r}{\tilde n}^{\LC}(r)^2\biggr]^{-1}
\int dr_1r_1^2 \int dr_2r_2^2
{\tilde n}^{\LC}(r_1){\tilde n}^{\LC}(r_2)D_1(\eta_0-r_1)D_1(\eta_0-r_2)
\nonumber \\
&& \times {4\over\pi^2}\int^{\infty}_0dk_1k_1^2P(k_1)\sum_{l}(2l+1)
\prod^2_{i=1}\biggl\{\biggl[b(k_1;\eta_0-r_i)-k_1^{-2}
{\cal D}_{r_i}\biggr]j_l(k_1r_i)\biggl\}
\nonumber \\
&&\times\int dk k^2 j_l(kr_1)j_l(kr_2)j_0(kR)~.
\label{eq60}
\end{eqnarray}
By using the mathematical formulae, (Magnus et~al. 1966)
\begin{eqnarray}
\int dk k^2 j_l(kr_1)j_l(kr_2)j_0(kR)=\left\{
\begin{array}{ll}
{\pi\over 4r_1r_2R}P_l\biggl({r_1^2+r_2^2-R^2 \over 2r_1r_2}\biggr)
& (|r_1-r_2|<R<r_1+r_2),
\\
0 & (R<|r_1-r_2|,R>r_1+r_2),
\end{array}\right.
\label{eq61}
\end{eqnarray}
and 
\begin{equation}
  \sum_l(2l+1)P_l(\cos \theta)j_l(kr_1)j_l(kr_2)
  =j_0\biggl(k\sqrt{r_1^2+r_2^2-2r_1r_2\cos \theta}\biggr),
\end{equation}
we find
\begin{eqnarray}
&&\xiLC(R)=\biggl[\int d^3\vect{r}{\tilde n}^{\LC}(r)^2\biggr]^{-1}
{1\over \pi R}
\int\!\!\!\int_{\cal S}dr_1dr_2r_1r_2{\tilde n}^{\LC}(r_1){\tilde n}^{\LC}(r_2)
D_1(\eta_0-r_1)D_1(\eta_0-r_2)
\nonumber \\
&& \hspace{1cm}\times \int^{\infty}_0dk k^2 P(k)
\prod^2_{i=1}\biggl[b(k;\eta_0-r_i)-k^{-2}{\cal D}_{r_i}\biggr]
j_0\biggl(k\sqrt{r_1^2+r_2^2-2r_1r_2\cos \theta}\biggr)~,
\label{eq64}
\end{eqnarray}
where ${\cal S}$ denotes the region $|r_1-r_2|\leq R \leq r_1+r_2$,
and $\cos\theta$ is replaced by $\cos\theta=(r_1^2+r_2^2-R^2)/2r_1r_2$
after operating the differentiations with respect to $r_1$ and $r_2$.
 
Using the formulae, (B17)-(B19) in Paper \II,
and omitting the second term in the derivative, i.e., approximating as
${\cal D}_r\simeq f(\eta_0-r)\partial^2/\partial r^2$, we finally have
\begin{eqnarray}
\xiLC(R)&=&\biggl[\int dr r^2{\tilde n}^{\LC}(r)^2\biggr]^{-1}
{1\over 2R}\int\!\!\!\int_{\cal S}dr_1dr_2r_1r_2
{\tilde n}^{\LC}(r_1){\tilde n}^{\LC}(r_2)
\nonumber \\
&&\times{1\over 2\pi^2}\int dk k^2 P(k)
\prod_{i=1}^2\biggl[{b(k;\eta_0-r_i)D_1(\eta_0-r_i)}\biggr]
\nonumber \\
&&\times\biggl[j_0(kR)+\beta(k;\eta_0-r_2)I(k;R;r_1,r_2)
+\beta(k;\eta_0-r_1)I(k;R;r_2,r_1)\biggr.
\nonumber \\
&&\biggl.
+\beta(k;\eta_0-r_1)\beta(k;\eta_0-r_2)J(k;R;r_1,r_2)\biggr],
\label{eq68}
\end{eqnarray}
where 
\begin{eqnarray}
I(k;R;r_1,r_2)&=&{j_1(kR)\over kR}-{j_2(kR)\over R^2}
\biggl\{{R^2+r_2^2-r_1^2\over 2r_2}\biggr\}^2
\label{eq69}
\\
J(k;R;r_1,r_2)&=&{j_2(kR)\over (kR)^2}\biggl[2\biggl\{
{r_1^2+r_2^2-R^2\over 2r_1r_2}\biggr\}^2+1\biggr]
\nonumber \\
&&+{j_4(kR)\over R^4}\biggl\{{R^2+r_1^2-r_2^2\over 2r_1}\biggr\}^2
\biggl\{{R^2+r_2^2-r_1^2\over 2r_2}\biggr\}^2
\nonumber \\
&&-{j_3(kR)\over kR^3}\biggl[\biggl\{{R^2+r_1^2-r_2^2\over 2r_1}\biggr\}^2
+\biggl\{{R^2+r_2^2-r_1^2\over 2r_2}\biggr\}^2\biggr.
\nonumber \\
&&\biggl.-{R^2+r_1^2-r_2^2\over r_1}{R^2+r_2^2-r_1^2\over r_2}
{r_1^2+r_2^2-R^2\over 2r_1r_2}\biggr],
\label{eq70}
\end{eqnarray}
and $\beta(k;\eta_0-r)$ is defined by equation (\ref{eq71}).
Equation (\ref{eq68}) is identical to equation (\ref{eq68.5}).

\newpage
\end{appendix}
\centerline{\bf REFERENCES}
\bigskip
\def\apjpap#1;#2;#3;#4; {\pp#1, {#2}, {#3}, #4}
\def\apjbook#1;#2;#3;#4; {\pp#1, {#2} (#3: #4)}
\def\apjppt#1;#2; {\pp#1, #2.}
\def\apjproc#1;#2;#3;#4;#5;#6; {\pp#1, {#2} #3, (#4: #5), #6}

\apjpap Alcock, C., \& Paczynski, B. 1979;Nature;281;358;
\apjpap Ballinger, W. E., Peacock, J. A., \& Heavens, A. F. 
  1996;MNRAS;282;877;
\apjpap Bardeen, J. M., Bond, J. R., Kaiser, N., \& Szalay,
A. S. 1986;ApJ;304;15;
\apjpap Cole, S., Fisher, K. B., \& Weinberg, D. H. 1995;MNRAS;275;515;
\apjpap Davis, M., \& Peebles, P. J. E. 1983;ApJ;267;465;
\apjpap Dekel, A., \& Lahav, O. 1999;ApJ;520;24;
\apjpap de Laix ,A. A., \& Starkman, G. D. 1998;MNRAS;299;977;
\apjpap Feldman, H. A., Kaiser, N., \& Peacock, A. A. 1994;ApJ;426;23;
\apjpap Fry, J. N. 1996;ApJ;461;L65;
\apjppt Hamilton, A. J. S. 1998; in `` The Evolving Universe. Selected
Topics on Large-Scale Structure and on the Properties of Galaxies'',
(Kluwer: Dordrecht), p.185;
\apjpap Kaiser, N. 1987;MNRAS;277;1;
\apjpap Kitayama, J., \& Suto, Y. 1997;ApJ;490;557;
\apjpap Kodama, H., \& Sasaki, M.  1984;Prog. Theor. Phys. Supp.;78;1;
\apjpap Loveday, J., Peterson, B. A., Efsathiou, G., \& Maddox, S.J.
1992;ApJ;390;338;
\apjppt Magira, H., Jing, Y. P., \& Suto, Y. 1999;ApJ, in press;
\apjbook Magnus, W., Oberhettinger, F., \& Soni, R. P. ;Formulae and 
Theorems for the Special Functions of Mathematical 
Physics, P.426;Springer-Verlag, Belrin;1966;
\apjpap Matarrese, S., Coles, P., Lucchin, F., \& Moscardini, L
1997;MNRAS;286;115;
\apjppt Matsubara, T. 1999;astro-ph/9908056;
\apjpap Matsubara, T., Suto, Y., \& Szapdi, I. 1997;ApJ;491;L1;
\apjpap Matsubara, T., \& Suto, Y. 1996;ApJ;470;L1;
\apjpap Moscardini, L., Coles, P., Lucchin, F., \& Matarrese, S. 1998;MNRAS;
        299;95;
\apjpap Nakamura, T. T., Matsubara, T., \& Suto, Y. 1998;ApJ;494;13;
\apjpap Nakamura, T. T., \& Suto, Y. 1997;Prog. Theor. Phys.;97;49;
\apjpap Nishioka, H., \& Yamamoto, K. 1999;ApJ;520;426~(Paper \II);
\apjpap Ryden, B. S. 1995;ApJ;452;25;
\apjpap Sugiyama, N. 1995;ApJS;100;281;
\apjpap Suto, Y., Magira, H., Jing, Y. P., Matsubara, T., \& Yamamoto, K.
1999;Prog. Theor. Phys. Supp.;133;183;
\apjppt Suto, Y., Magira, H., \& Yamamoto, K. 1999;submitted to PASJ;
\apjpap Szalay, A.S., Matsubara, T., \& Landy, S.D. 1998;ApJ;498;L1;
\apjpap Taruya, A., Koyama, K., \& Soda, J. 1999;ApJ;510;541;
\apjpap Wallington, S., \& Narayan, R. 1993;ApJ;403;517;
\apjpap Yamamoto, K., \& Suto, Y. 1999;ApJ;517;1~(Paper I);
\apjppt Yamamoto, K., Nishioka, H., \& Suto, Y. 1999;ApJ 
in press~(Paper \III);  

\newpage
\newpage
\newpage
\newpage
\newpage
\newpage
\end{document}